# Global sensitivity analysis based on DIRECT-KG-HDMR and thermal optimization of pin-fin heat sink for the platform inertial navigation system

Xin Luo[a], Fengtao Xu[b], Ting Wang[b], Hu Wang[a*], Yang Zeng[a]

[a]*State Key Laboratory of Advanced Design and Manufacturing for Vehicle Body, Hunan University, Changsha 410082, P.R. China*

[b]*Beijing Institute of Aerospace Control Devices, Beijing 100854, P.R.China*

**Abstract**: In this study, in order to reduce the local high temperature of the platform in inertial navigation system (PINS), a pin-fin heat sink with staggered arrangement is designed. To reduce the dimension of the inputs and improve the efficiency of optimization, a feasible global sensitivity analysis (GSA) based on Kriging-High Dimensional Model Representation with DIviding RECTangles sampling strategy (DIRECT-KG-HDMR) is proposed. Compared with other GSA methods, the proposed method can indicate the effects of the structural and the material parameters on the maximum temperature at the bottom of the heat sink by using both sensitivity and coupling coefficients. From the results of GSA, it can be found that the structural parameters have greater effects on thermal performance than the material ones. Moreover, the coupling intensities between the structural and material parameters are weak. Therefore, the structural parameters are selected to optimize the thermal performance of the heat sink, and several popular optimization algorithms such as GA, DE, TLBO, PSO and EGO are used for the optimization. Moreover, steady thermal response of the PINS with the optimized heat sink is also studied, and its result shows that the maximum temperature of high temperature region of the platform is reduced by 1.09 ℃ compared with the PINS without the heat sink.

***Keywords:*** Pin-fin heat sink; Steady thermal response; Global sensitivity analysis; DIRECT-KG-HDMR; Thermal optimization.

*Corresponding author, email: wanghu@hnu.edu.cn, Tel: 86-731-88821445, Fax: 86-731-88822051

**Nomenclature**

| | | | | |
|---|---|---|---|---|
| $c_1$ | personal learning coefficient | | $T$ | temperature (℃) |
| $c_2$ | global learning coefficient | | $v$ | velocity (*m/s*) |
| $C_p$ | heat capacity (*J/kg K*) | | $W$ | width of computational domain (*mm*) |
| $d$ | distance (*mm*) | | | |
| $h_d$ | height of base plate (*mm*) | | *Greek symbols* | |
| $h_f$ | height of pin fin (*mm*) | | $\rho$ | density ($kg/m^3$) |
| $F$ | scaling factor | | $\theta$ | phase angle of the polygon (°) |
| $H$ | height of computational domain (*mm*) | | $\mu$ | dynamic viscosity (*Pa s*) |
| $k$ | thermal conductivity (*W/m K*) | | $\omega$ | inertia weight |
| $L$ | length of computational domain (*mm*) | | | |
| $m_l$ | number of fins at odd-numbered row in length | | *Subscripts* | |
| $m_w$ | number of fins at odd-numbered row in width | | *ave* | average |
| *Maxit* | maximum allowed iteration | | *bottom* | at bottom of the heat sink |
| $n$ | number of regular polygon edges | | *f* | fluid |
| *nM* | number of mutants | | *in* | inlet |
| *nPop* | population size | | *l* | in length |
| $p$ | pressure (*Pa*) | | *max* | maximum |
| *pC* | crossover rate | | *min* | minimum |
| *pM* | mutation rate | | *out* | outlet |
| $q$ | heat flux ($W/m^2$) | | *s* | solid |
| $r$ | radius of regular *n*-sided polygon (*mm*) | | *w* | in width |

# 1. Introduction

Currently, PINS has been a widely used system to maintain a moving object such as aircraft, ship in a specific orientation in space [1]. It is very complicated and consists of computer, accelerometers, gyroscopes, platform and some other important devices. The platform is a carrier of the gyroscopes, accelerators and other important components. The temperature distribution of the platform has a very large influence on the accuracies of gyroscopes and accelerometers. If imprecise measurements are used, it would lead to disastrous accident.

In order to improve the temperature stability of the inertial devices, Ma et al. [2] proposed a Fuzzy Proportional Integral Derivative (Fuzzy-PID) control method. He took fuzzy adjustment to the PID parameters and proved that this method could meet the accuracy requirements of quick starting and output stability with a well-designed experiment. Zhou et al. [3] investigated a temperature control system based on the integrated temperature measurement digital circuit. In his study, with this system, the temperature controlling precision could reach ±0.4 ℃ when the temperature variation was limited in the range of ±1 ℃. Huang et al. [4] performed another work to compare the temperature control performance of INS between Inner Model Control (IMC) and PID. They found that the control performance of IMC was better than PID. An algorithm based on single neuron adaptive PID control was presented in Ref. [5]. In this study, the values of corresponding proportion, integral and differential coefficients can be adjusted automatically by the learning ability of neurons. Ren et al. [6] carried out an applied temperature control method to analyze the relationship between the environmental temperature and measurement accuracy of gyroscopes. Martorana et al. [7] tried to control the temperature of the gyroscopes and accelerometers by Active Bimodal Control (ABC) method. He proved that the ABC was an efficient way to steady the temperature fluctuation within ±0.1 ℃. Grigorie et al. [8] presented a new method to recognize the correction between the variety of temperature and the gyroscope's bias variation, which was obtained with classical compensation algorithm by a fuzzy logic controller. Moreover, the fuzzy logic controller was developed via a fuzzy inference system (FIS) based on fuzzy neural network (FNN). It verified that the method was more accurate than the classical algorithms based on the Least Squares Method (LSM). Liu et al. [9] concluded that the main factors, which could lead to changes of temperature, are the variety of working temperature and the radiation resulting from the Fiber Optic Gyroscope (FOG). They also found an Automatic Temperature Control (ATC) system could steady the temperature fluctuation within ±0.1 ℃ when the ambience temperature varied from -30 ℃ to 30 ℃. Wu et al. [10] found that the temperature was the primary factor to influence the output of the gyroscopes. They experimentally investigated the bias of gyroscopes that could be compensated mostly by temperature, temperature gradient and temperature rate. Dzhashitov et al. [11] carried out a research on nanostructure composite materials based on carbon nanotubes with high thermal conductivity, which could be applied in inertial devices to reduce the temperature variance. Fontanella et al. [12] presented an innovative calibration method based on the use of Back-Propagation Neural Networks (BPNNs). The purpose of this work was to calibrate the surrogate model, which indicated the correlation between Micro-Electro-Mechanical System (MEMS) gyroscope null-voltage and temperature. They also compared the adaptive calibration method to the traditional method and summarized the advantages of the proposed method. Golikov et al. [13] proposed a new method to compensate the multiplicative and additive errors caused by temperature changes. Wang et al. [14] presented a new method to compensate the error and improved the precision of Fiber-Optic Gyroscopes (FOGs) based on improved Particle Swarm Optimization (PSO) and Support Vector Machine (SVM) algorithms. The regression accuracy of the proposed method increased by 83.81% compared with the traditional SVM. Considering the electromechanical-thermal and the dynamic thermal errors caused by accelerations, the integrated thermal error compensation method was proposed and the bias temperature sensitivity could be reduced by more than one order of magnitude compared with the raw bias temperature sensitivity of the gyroscope [15]. A fault-tolerant navigation system configuring with six independent sensor channels, which consisted of the paired gyroscopes and accelerometers, has been developed. The high navigation accuracy of the new proposed system was achieved when the gyroscopes operated over a wide temperature range [16]. Considering the limitations of Least Squares fitting (LSF) and Neural Network (NN), Wei et al. [17] proposed a new scheme of system-level temperature compensation of RLG bias based on Least Squares-Support Vector Machine (LS-SVM). They also experimentally investigated the temperature of Ring Laser Gyroscope (RLG) bias to validate the effectiveness of the proposed method. Generally, these methods can be classed as three types, temperature control, temperature compensation and usage of the low temperature-sensitive materials for the platform. Temperature control is to keep the operating temperature of the gyroscopes and inertial measurement system as uniformly distributed as possible. Furthermore, the performances of these methods are related to the locations of the temperature-measuring points. As for the temperature compensation, the temperature can be compensated by the gyroscope output signal-processing algorithm. However, it is not a convenient approach for different systems because the compensation methods are different to be carried out due to different temperature characteristics. Moreover, the temperature difference of the platform can be controlled via the low temperature-sensitive materials, such as Phase Change Material (PCM). Compared with the first two methods, the most difficulty is to improve the thermal property of material in a relatively short period.

In this work, the heating film is used in the inner surfaces of the gyroscope-mounting holes to reach a given temperature quickly before working. However, it would result in non-uniform temperature distribution of platform and affect the temperature stability of the gyroscopes which are installed in the holes. Thus, an alternative way to reduce the temperature difference of the platform which caused by the uneven heating in the region where heat sources are concentrated and to improve the temperature stability of gyroscopes is needed. Heat sink has been identified as one of the most promising and effective cooling methods due to its excellent performance and high reliability [18]. Therefore, a heat sink mounted on the high temperature region of the platform would be a good way to improve temperature stability of the platform. In order to improve the thermal stabilization of inertial measurement units (IMU), the heat sink was compressed together with printed circuit board (PBC) and thermoelectric devices [19]. It should be noted that the thermal performance of the heat sink is mainly related to shape and arrangement of the pin fin. Joo et al. [20] proposed a new correlation of the heat transfer coefficient to optimize the heat sink. In their opinion, heat dissipated by the optimized pin-fin heat sinks was more than heat dissipated by the plate-fin heat sinks under the same conditions. Mon et al. [21] investigated the effects of fin spacing on the performances of annular-finned tube heat exchangers in staggered and in-line arrangements. They found that fin spacing to height ratio and Reynolds number have a great influence on the boundary layer development. Maji et al. [22] compared the thermal performance of heat sink using perforated pin fins with linear and staggered arrangements, they found that the thermal performance of the heat sink with staggered arrangement was better than the linear one. Ho et al. [23] investigated the airfoil shaped fins with staggered arrays and the result was compared with the circular and rounded rectangular fins. They found that the thermal performances of the airfoil and rounded rectangular heat sink were better than those of the circular heat sink. Xia et al. [24] investigated the influence of several parameters on the total thermal resistance and pumping power based on Multi-Objective Evolutionary Algorithm (MOEA). They revealed that higher rib would lead to a higher thermal resistance and pumping power. The thermal performance of ten various configurations of heat sink were also studied in Ref [25]. The results showed that the thermal performance was not improved when the fin space was less than 3 *mm*. Rao et al. [26] optimized the entropy generation rate and material cost of the heat sink based on Teaching-Learning-based Optimization (TLBO). Based on topology optimization, the heat sink with different structure were obtained according to the input power and the heat source conditions in Ref [27]. Moreover, a new approach called SFCVT for single response adaptive design of deterministic computer experiments was presented and applied in optimizing the heat exchanger to get more compact structure [28].The material of the heat sink is another key factor in capability of heat transfer. Kuo et al. [29] experimentally investigated that the thermal performance of the heat sink consisting of aluminum-graphite pin fins improved with the increase of the content of graphite.

In this study, in order to improve the thermal performance of the platform, a heat sink using pin fins with staggered arrangement is designed. Moreover, the DIRECT-KG-HDMR is proposed to perform the global sensitivity analysis. The structural parameters $h_f$, $h_d$, $r$, $m_w$, $m_l$, $n$, $\theta$ and the material parameters $\rho$, $C_P$, $k$ are selected to indicate the correlations between design variables and response function. Sensitivity and coupling coefficients are defined and discussed. Then, the insensitive material parameters are filtered. An optimization design is implemented based on the selected sensitive structural parameters. Several popular optimization algorithms such as GA, DE, TLBO, PSO and EGO, are used to optimize the thermal performance of the heat sink and the best one of these optimization solutions is used for design. Finally, steady thermal response of the PINS with the optimized heat sink is studied to validate the effectiveness of the heat sink.

# 2. Problem descriptions

## 2.1. Background

Local high temperature of the platform which is caused by the uneven distribution of heat sources is the main reason of the large temperature difference in the platform. The schematic diagram of PINS is shown in Fig. 1, and a 3D model of the platform with pin-fin heat sink is shown in Fig. 2 in which heat sources concentrated near Face 1 is important due to the local high temperature of the platform. Therefore, a heat sink using pin fins with staggered arrangement mounted on Face 1 is designed to reduce the temperature difference of the platform via reducing the local high temperature of the platform.

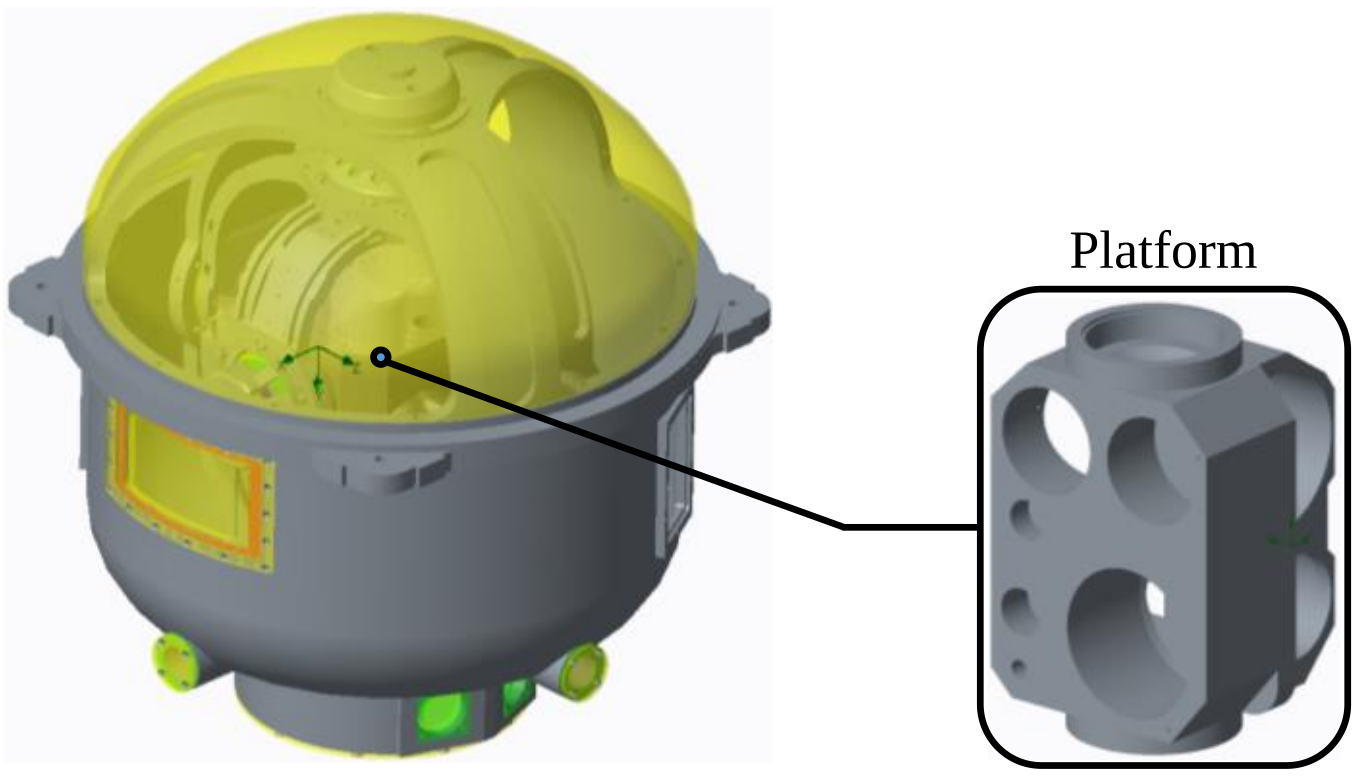

**Fig. 1.** Schematic diagram of PINS.

## 2.2. Physical model

The heat sink, which is composed of the base plate and pin fins, is presented in Fig. 3. Heat transfer improves when fluid flows in the curved channels are created by pin fin arrays in staggered arrangement. Considering the platform structure, the base plate of heat sink is 70 *mm* in length, 30 *mm* in width and $h_d$ in height. The height of the fin is $h_f$, and the cross-sectional shape of the fin is a regular *n*-sided polygon with circumscribed circle radius *r*. The phase angle of the polygon is $\theta$. There are $m_l$ pieces of fins at odd-numbered row and $(m_l - 1)$ pieces of fins at even row in width. In a word, the number of fins is $(2m_w - 1)$ in width. Therefore, the distance between the circumscribed circles of the neighbor fins in length is:

$$d_l = (70 - 2m_l r)/m_l \tag{1}$$

The distance between the circumscribed circles of the neighbor fins in width is:

$$d_w = [30 - 2(2m_w - 1)r]/(2m_w - 1) \tag{2}$$

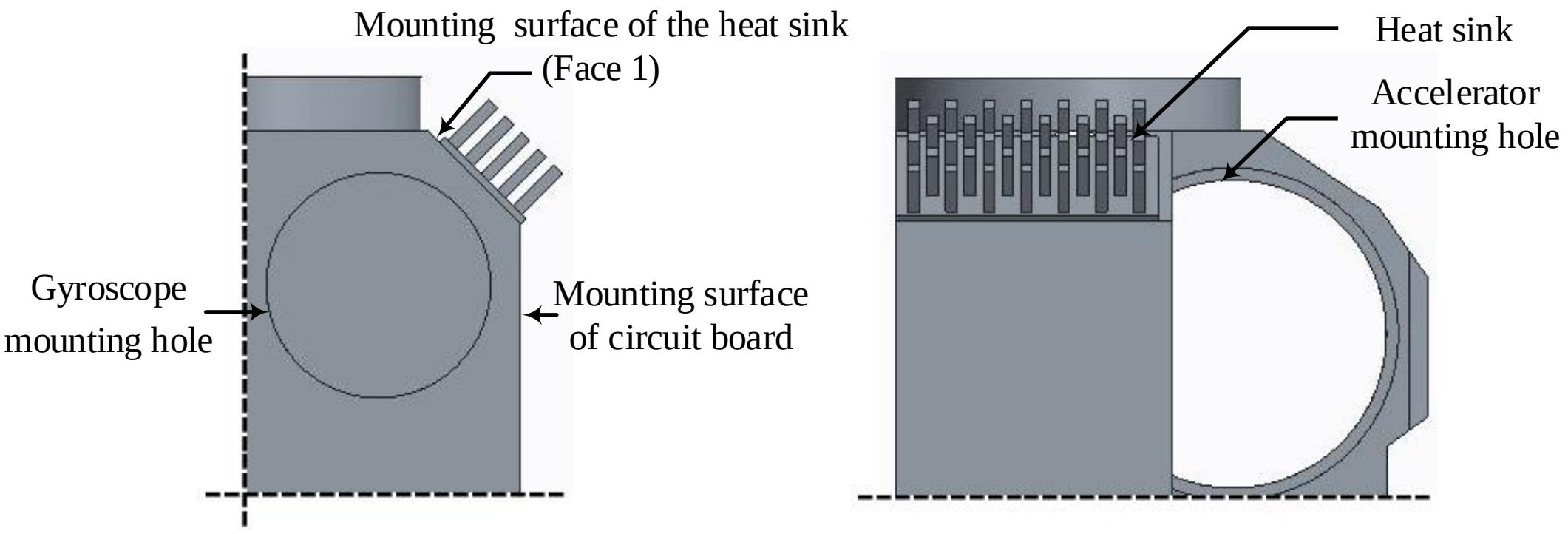

(a) Top view (b) Right view

**Fig. 2.** Schematic diagram of the platform with heat sink.

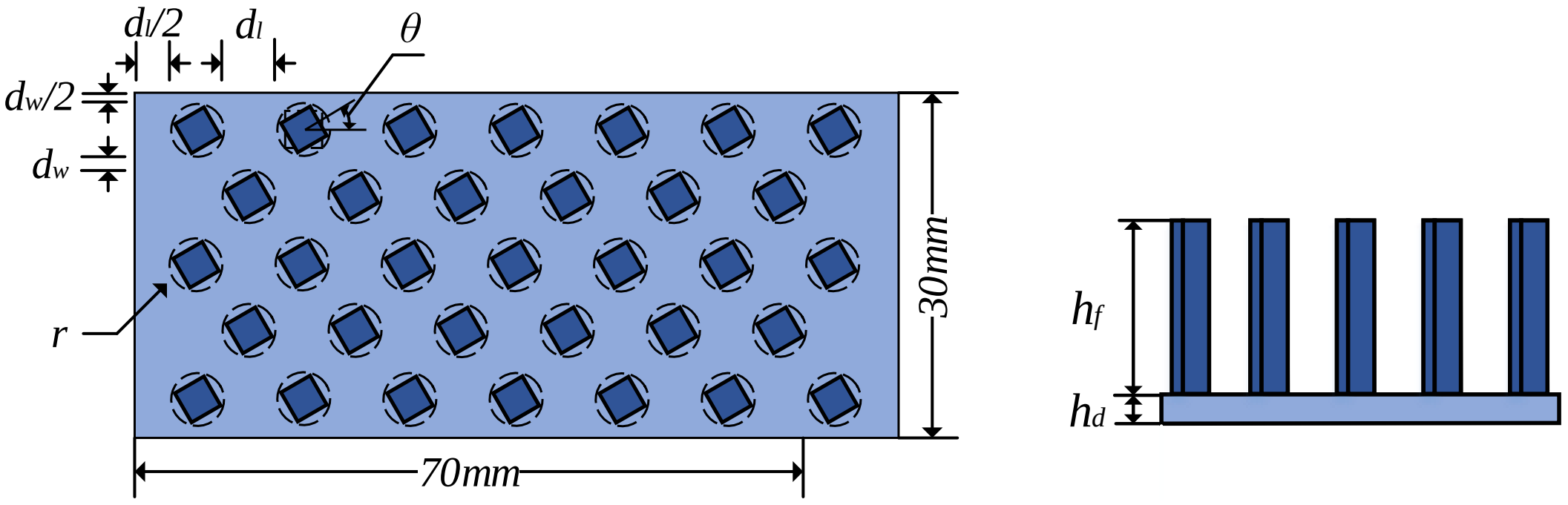

(a) Top view (b) Left view

**Fig. 3.** The geometry of heat sink.

Air is used as the coolant because of the working environment of PINS, and properties of air are listed in 0 [30]. The heat sink is made of aluminum alloy. The density, heat capacity, thermal conductivity of aluminum alloy are 2700 $kg/m^3$, 890 $J/(kg\ K)$, 155 $W/(m\ K)$, respectively.

**Table 1.** Properties of air.

| Temperature ($K$) | $\rho$ ($kg/m^3$) | $C_p$ ($J/(kg\ K)$) | $k$ ($W/(m\ K)$) | $\mu$ ($Pa\ s$) |
|---|---|---|---|---|
| 300 | 1.177 | 1004.9 | 0.02624 | $1.85\times10^{-5}$ |
| 325 | 1.085 | 1006.3 | 0.02816 | $1.96\times10^{-5}$ |
| 350 | 1.007 | 1008.2 | 0.03003 | $2.10\times10^{-5}$ |

## 2.3. Mathematical model

The following assumptions are made to build the mathematical model:

1) Fluid flow is incompressible, laminar, and steady-state.
2) The thermo-physical properties of aluminum alloy are temperature independent.
3) No velocity slip at the walls.

Based on the above assumptions, the governing equations are expressed as follows:

*Continuity equation* is the specific form of mass conservation law in fluid mechanics and it is expressed as:

$$\vec{\nabla}\cdot\vec{v} = 0 \tag{3}$$

where $\vec{v}$ is the flow velocity.

*Momentum equation* is expressed as:

$$\rho_f(\vec{v}\cdot\vec{\nabla})\vec{v} = -\vec{\nabla}P + \mu_f\nabla^2\vec{v} \tag{4}$$

where $\rho_f$ and $\mu_f$ are the density and viscosity of the fluid, respectively.

*Energy equation* is expressed as:

$$\rho_f C_{p,f}(\vec{v}\cdot\vec{\nabla}T_f) = k_f\nabla^2 T_f \tag{5}$$

$$0 = \nabla^2 T_s \tag{6}$$

where $C_{p,f}$ and $k_f$ are the specific heat capacity and thermal conductivity of the fluid, respectively.

## 2.4. Validation of the PINS

The experiment setup of the PINS is shown in Fig. 4. Firstly, the velocity of the fan and power of the heating film are controlled by the controllable fan velocity unit and electronic control unit, respectively. Meanwhile, the platform is equipped with seven T-type thermocouples to measure the temperature of eight temperature-measuring points during the experimental work; the dots shown in Fig. 5 indicate the locations of the thermocouples. Then, the signals of the thermocouples should be transmitted to the data acquisition system, and then, it would be converted and transmitted to the host computer for further operation.

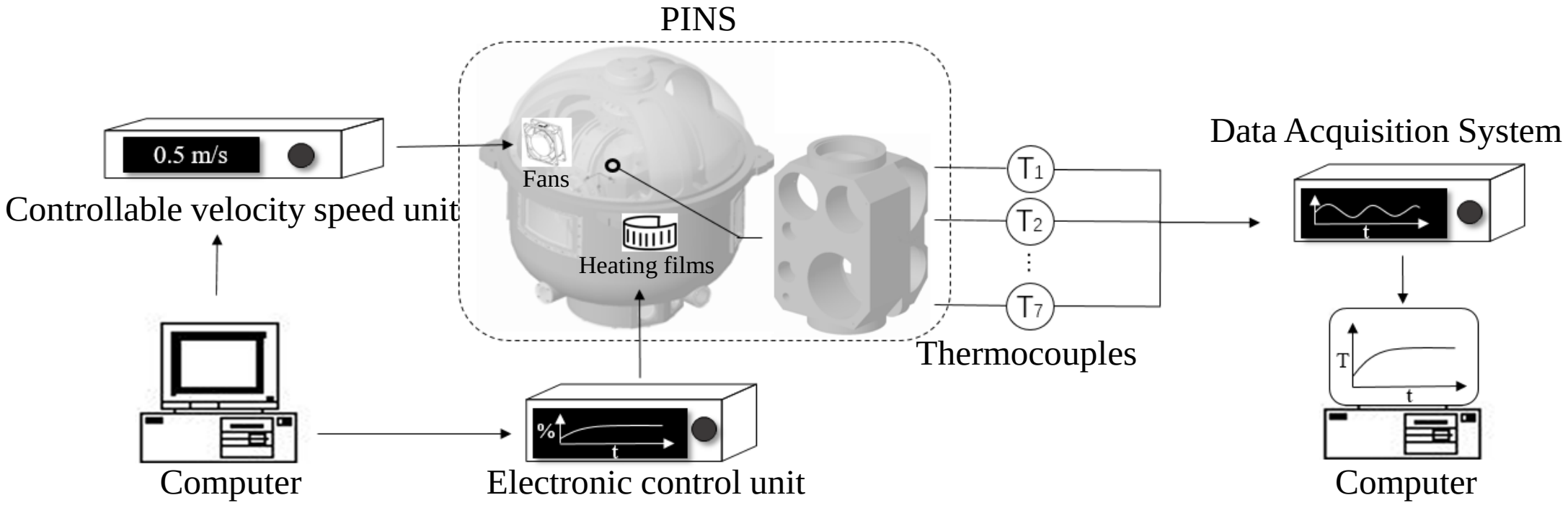

**Fig. 4.** Experiment setup.

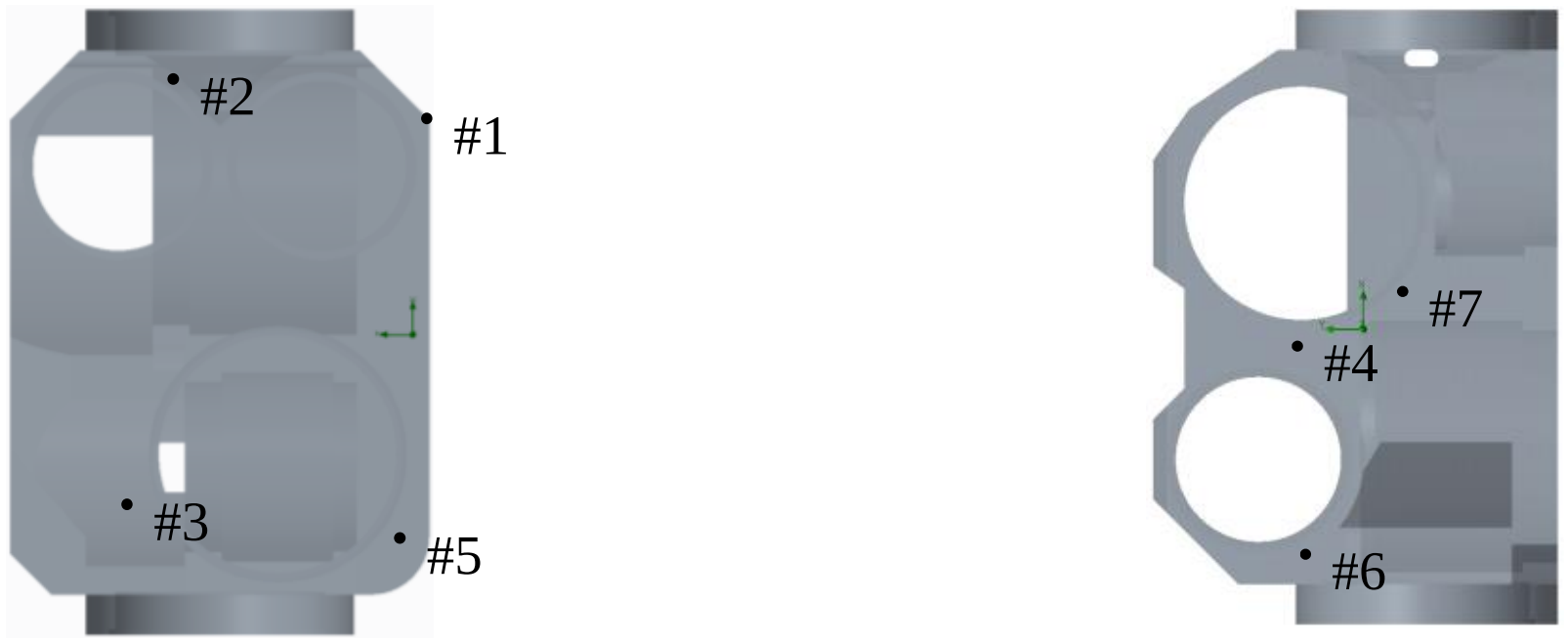

(a) Top view (b) Left view

**Fig. 5.** Locations of the thermocouples.

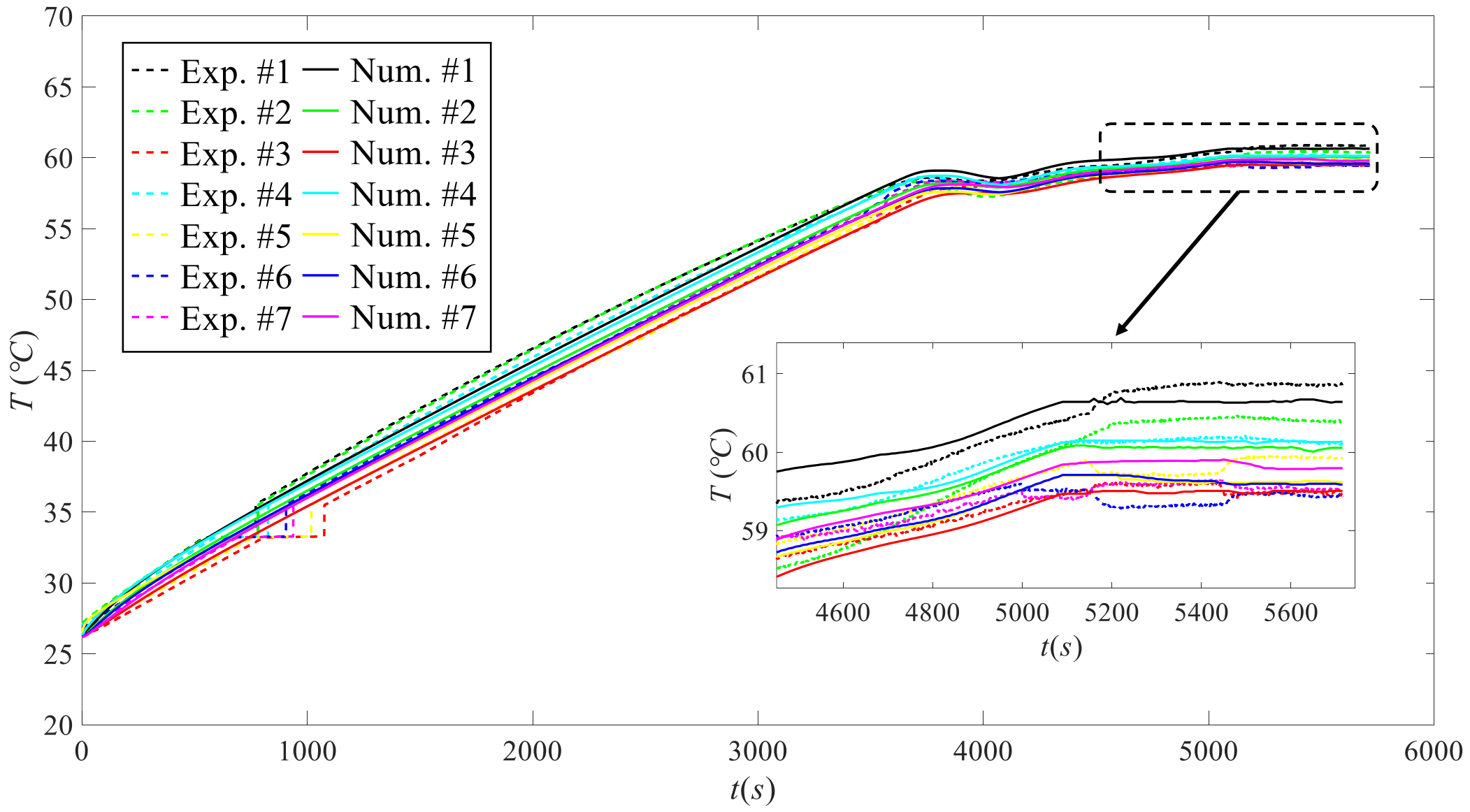

**Fig. 6.** Comparison between the simulation and experimental results.

To assess the accuracy of the simulation setup for the PINS in this work, the comparison between the simulation and experiment results of the temperature at different temperature-measuring points is shown in Fig. 6. The experimental result is indicated by a solid line and simulation result by a dash line. During the heating stage, the temperature rises rapidly in order to reach the working temperature as quickly as possible. Then, the temperature reaches a steady-stage at the working stage. It could be observed that the simulation result is in well matched with the experimental result.

# 3. GSA based on DIRECT-KG-HDMR

In this study, GSA is introduced to analyze the effects of the structural parameters and material parameters on the thermal performance of the pin-fin heat sink with staggered arrangement.

## 3.1. Preparation for sensitivity analysis

Sensitivity analysis is a method that reveals how the uncertainty of the inputs affects the uncertainty of the output [31]. The inputs that are not sensitive to the output will not be considered in thermal optimization. Sequentially, the dimension of the inputs can be reduced and the efficiency of optimization can be improved. Currently, sensitivity analysis can be divided into Local Sensitivity Analysis (LSA) and GSA. Due to the advantage of high efficiency, the LSA has been widely used in many fields. Specifically, One -at-a-time (OAT) [32] is the most common method of LSA. It was used to identify the inputs that are most influential on the thermal behavior of the test cell [33]. However, it does not fully explore the input space and cannot indicate the correlations between the inputs. Therefore, the GSA is a more popular way to identify the complex nonlinear models because it can analyze the influence of the interaction between the inputs and the output, and identfy the most influential inputs under the condition of *isonecitance equivalent*. At present, the commonly used GSA are Variance-based sensitivity analysis [34], Fourier Amplitude Sensitivity Test (FAST) [35], HDMR [36] and so on.

HDMR introduced by Rabitz et al. aims to quantitatively analyze and evaluate the relationship between inputs and the response function in this study. The HDMR provides a limited-level and interrelated function expansion structure that can decompose a high-dimensional problem into the sum of multiple low-dimensional problems. Each level of the function is constructed based on the function of the previous level, and the same level of components can be solved independently. Specifically, $f(\mathbf{x})$ is the corresponding response function of the inputs in the design space of the $n$-dimensional vector $\mathbf{x} = [x_1, x_2, \dots, x_n]^T \in R^n$, it is expressed as a finite expansion for a given multivariable response function in terms of input variables and its general form is [37] :

$$f(\mathbf{x}) = f_0 + \sum_{i=1}^{n} f_i(x_i) + \sum_{1<i<j\leq n} f_{ij}(x_i, x_j) + \sum_{1<i<j<k\leq n} f_{ijk}(x_i, x_j, x_k)$$
$$+\sum_{1<i_1<\cdots<i_l\leq n} f_{i_1 i_2 \dots i_l}(x_{i_1}, x_{i_2}, \dots, x_{i_l}) + \cdots + f_{1,2,\dots,n}(x_i, x_j, \dots, x_n) \quad (7)$$

where $n$ denotes the number of input variables of an underlying problem. $f_0$ is a constant representing the value at the center of the function, $f_i(x_i)$ is the first-order term that indicates the influence of the variable $x_i$ on the response function when it acts alone. $f_{ij}(x_i, x_j)$ is the second-order term that indicates the influence of the variables $x_i$ and $x_j$ on the response function, and so on. The last term indicates the effect on the response function when all inputs are coupled together.

In order to overcome the difficulties of high-dimensional modeling, KG-HDMR, which integrates Kriging and HDMR, is used to construct the approximate model of an underlying problem. Kriging, which was proposed by Krige in 1951, is the method of interpolation deriving based on regionalized variable theory [38]. It was originally used to predict the distribution of mineral resources. Some scholars extended Kriging to the field of numerical approximation for interpolation analysis of data. The mathematical form of Kriging is:

$$y(\mathbf{x}) = f^T(\mathbf{x})\boldsymbol{\beta} + z(\mathbf{x}) \quad (8)$$

where $f^T(\mathbf{x})\boldsymbol{\beta}$ is the global regression model of the actual response, and $\boldsymbol{\beta} = [\beta_1, \beta_2, \dots, \beta_p]^T$ is the regression coefficient, $f(\mathbf{x}) = [f_1(x), f_2(x), \dots, f_p(x)]^T$ is the pre-determined basis function vector. $z(\mathbf{x})$ is a random function of deviation that obeys normal distribution and satisfies the statistical properties as follow:

$$E\,[z(\mathbf{x})] = 0$$
$$Var\,[z(\mathbf{x})] = \sigma^2$$
$$Cov[z(\mathbf{x}^i), z(\mathbf{x}^j)] = \sigma^2 R(\mathbf{x}^i, \mathbf{x}^j) \quad (9)$$

where $\sigma^2$ is a scalar parameter called process variance, $R(\mathbf{x}^i, \mathbf{x}^j)$ is the binary correlation function that depends on the relative position of $\mathbf{x}^i$ and $\mathbf{x}^j$. Several correlation functions are used to represent $R(\mathbf{x}^i, \mathbf{x}^j)$. For example, if the power-exponential function for the correlation kernel is chosen, Equation 9 can be rewritten as:

$$Cov[z(\mathbf{x}^i), z(\mathbf{x}^j)] = \sigma^2 \cdot exp[-\sum_{k=1}^{n} \theta_k(|x_k^i - x_k^j|)] \quad (10)$$

where $\theta_k = [\theta_1, \theta_2, \ldots, \theta_k]$ is correlation parameters.

For complex engineering problems, the computational cost of the response function is extremely time-consuming. Consequently, intelligent sampling algorithm has become a research hotspot in the field of optimization. Such algorithms can automatically adjust the direction of the distribution according to the previous samples based on the feedback principle, and gradually converge to the neighborhood of the optimal solution. The performance of the sampling algorithm is a key factor in determining the quality and speed of the registration, and the ideal sampling algorithm should converge to the global optimal solution as quickly as possible. Since the search path for each iteration is independent of the optimal solution position obtained from the previous iteration, the DIRECT searches all regions that may contain global optimal solutions. Even if an iteration falls into a local optimum, it can guarantee to overstep the local optimum in the global scope. It can also preserve the region containing the global optimal solution in the iterative process and gradually refine it. Finally, the DIRECT can guarantee to find the global maximum of mutual information, which can completely avoid the influence of local optimum. Therefore, the DIRECT sampling strategy introduced by Jones et al. [39] is used in this study. DIRECT, which is effective for both local search and global search, is widely used for global optimization with boundary constrains. Fig. 7 is the schematic diagram of DIRECT sampling method. Firstly, the DIRECT normalizes design space to unit hypercube. Then, the unit hypercube can be divided into three sub-cube to find the potential optimal hypercube (Yellow circle in Fig. 7) based on the specific criteria. Sequentially, the potential optimal hypercube continues to be divided into three equal parts, and so on.

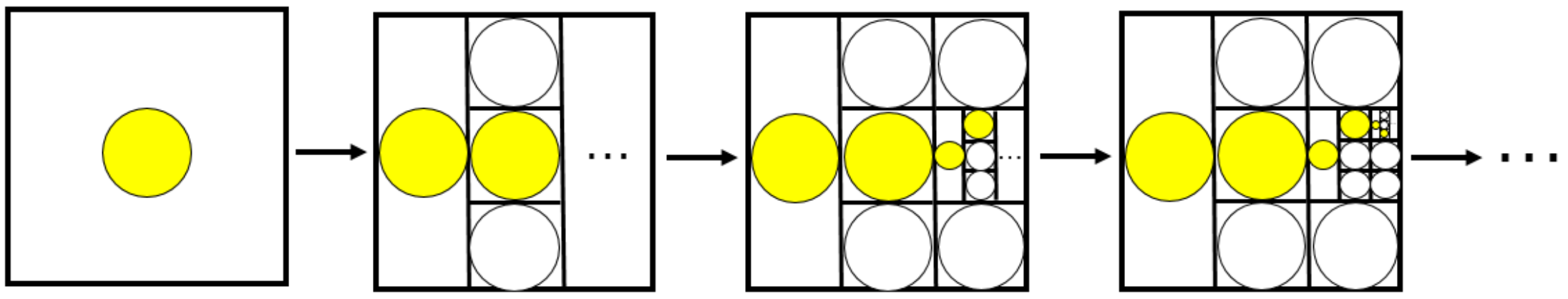

**Fig. 7.** Schematic diagram of DIRECT sampling method.

## 3.2. DIRECT-KG-HDMR

The KG-HDMR was originally proposed by Wang et al. [40] and has been applied to sheet forming optimization. DIRECT with Kriging approximation model was employed by Siah et al. [41] and has been used to solve optimization problems in the electromagnetic field. In this study, the adaptive meta-model technique DIRECT-KG-HDMR for multi-variable problems, which integrates HDMR, KG and DIRECT sampling method, is proposed to improve the efficiency and the accuracy of modeling procedure. It is well known that the uncoupled and low-order coupling terms are more sensitive to the response function practically. Therefore, only the uncoupled and the first-order coupling terms are considered in this study. Thus, the mathematical form of the DIRECT-KG-HDMR model is:

$$f(\mathbf{x}) \approx f_0 + \sum_{i=1}^{n}\left({f^{i}}^{T}(x_i)\boldsymbol{\beta}^i + z^i(\mathbf{x}_i)\right) + \sum_{1<i<j\ll n}\left({f^{ij}}^{T}(x_i, x_j)\boldsymbol{\beta}^{ij} + z^{ij}(x_i, x_j)\right) \quad (11)$$

The details of the DIRECT-KG-HDMR are described as follows.

*Step 1.* Choose the central point of the design space as the original vector $\mathbf{x} = [x_1, x_2, \dots, x_n]$ and calculate $f_0$.

*Step 2.* Construct the first-order component function $f_i(\mathbf{x}_i) = f([x_{10}, x_{20}, \dots, x_i, \dots, x_{n0}) - f_0$ by sampling in two ends of $x_i(x_{il}$ and $x_{iu})$ while fixing the rest of $x_j = x_{j0}\ (j \neq i)$ and the initial $\hat{f}_i(\mathbf{x}_i)$ is obtained.

*Step 3.* Is $f_i(x_i)$ linear or not? If $\left|(\hat{f}(\mathbf{x}_i) - f_0)/f_0\right| \leq 10^{-2}$ is satisfied, $f_i(\mathbf{x}_i)$ is assumed to be linear. Otherwise, continue to sample by the DIRECT and construct $f_i(\mathbf{x}_i)$ until the convergence criterion is satisfied.

*Step 4.* Loop through *step 2* and *step 3* until all the uncoupled terms have been constructed.

*Step 5.* Is the first-order coupling terms exists? Theoretically, the function value at each sample point in KG is accurate, in other words, $f_0 + \sum_{i=1}^{n} f_i\,(\mathbf{x}_i) = f_0 + \sum_{i=1}^{n} \hat{f}_i\,(\mathbf{x}_i)$ is satisfied in the set of sample points. Therefore, a new set of sample points $\mathbf{x}_{\boldsymbol{e}} = [x_{1e}, x_{2e}, \dots, x_{ie}, \dots, x_{je}, \ \dots, x_{ne}]$ is generated. For the general principle, one of the sample points $x_{il}$ and $x_{iu}$ is randomly taken as the *n*-th dimensional component $x_{ei}$ of the new sample. Within the accuracy range of the allowable error. If $f(\mathbf{x}_e) = f_0 + \sum_{i=1}^{n} \hat{f}_i\,(x_{ei})$ is satisfied, it can be considered that the first-order coupling terms do not exist and the procedure is terminated. Otherwise, it goes to the next step.

*Step 6.* Determine the coupled variable combinations. Generated a new set of sample points $[x_1, x_2, \dots, x_{ei}, \dots, x_{ej} \dots, x_n]^T$ $(1 \le i < j \le n)$ within the accuracy range of the allowable error. If $f\left(x_{ei}, x_{ej}, \mathbf{x}^{ij}\right) = f_0 + \hat{f}(x_{ei}) + \hat{f}\left(x_{ej}\right)$, it can be considered that the coupling term of $x_i$ and $x_j$ is invalid for the response function, it means that $x_i$ and $x_j$ are not coupled. Otherwise, the first-order coupling term exists and constructs based on KG. The process terminates when all the variable combinations have been identified.

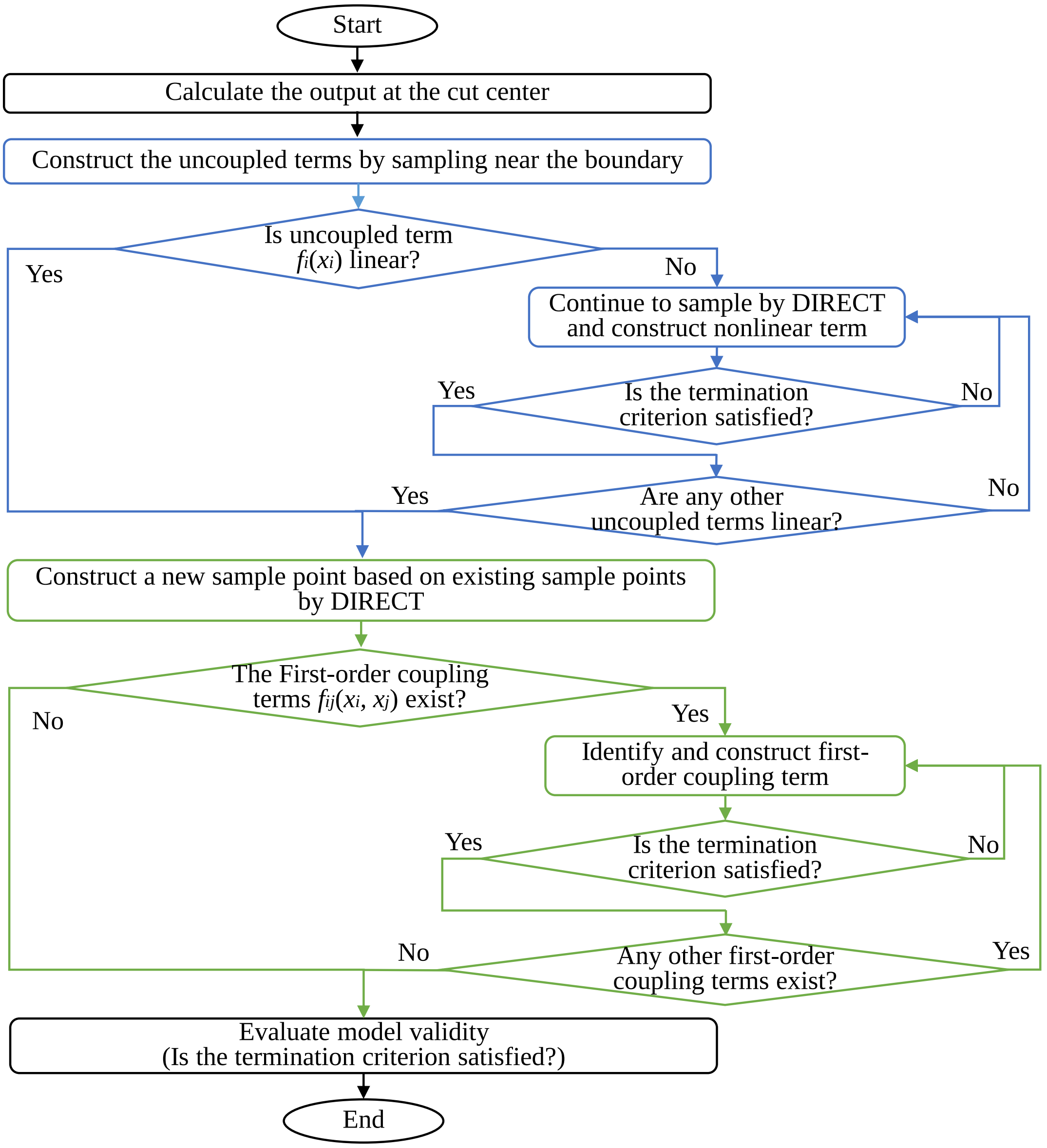

**Fig. 8.** Flowchart of DIRECT-KG-HDMR.

According the above mentioned suggested method, sampling strategy makes good use of the hierarchy of the HDMR, and it could reduce the difficulty of high-dimensional modeling by dividing the problem domain into multiple sub-domains. Moreover, it also discloses the characteristics of design variances (linear or non-linear, coupling or non-coupled) and the global characteristics of the response function by sampling points.

In addition, accuracy and convergence criteria are both defined by relative error. Accuracy criterion is used to identify coupling between input variances and its relative error is set to $10^{-5}$. Considering the computational efficiency of engineering problems, the relative error of convergence criterion is set to 0.1 in this study.

# 4. Optimization of pin-fin heat sink for PINS

## 4.1. Implementation of parameterized FE model

The numerical model of the heat sink is simplified as shown in Fig. 9. The computational domain of this model is $160\ mm\ (L) \times 70\ mm\ (W) \times 50\ mm\ (H)$, and the heat sink is located 50 $mm$ to the right of the velocity inlet. The boundary conditions are depicted in Fig. 9. The velocity-inlet is applied at the inlet:

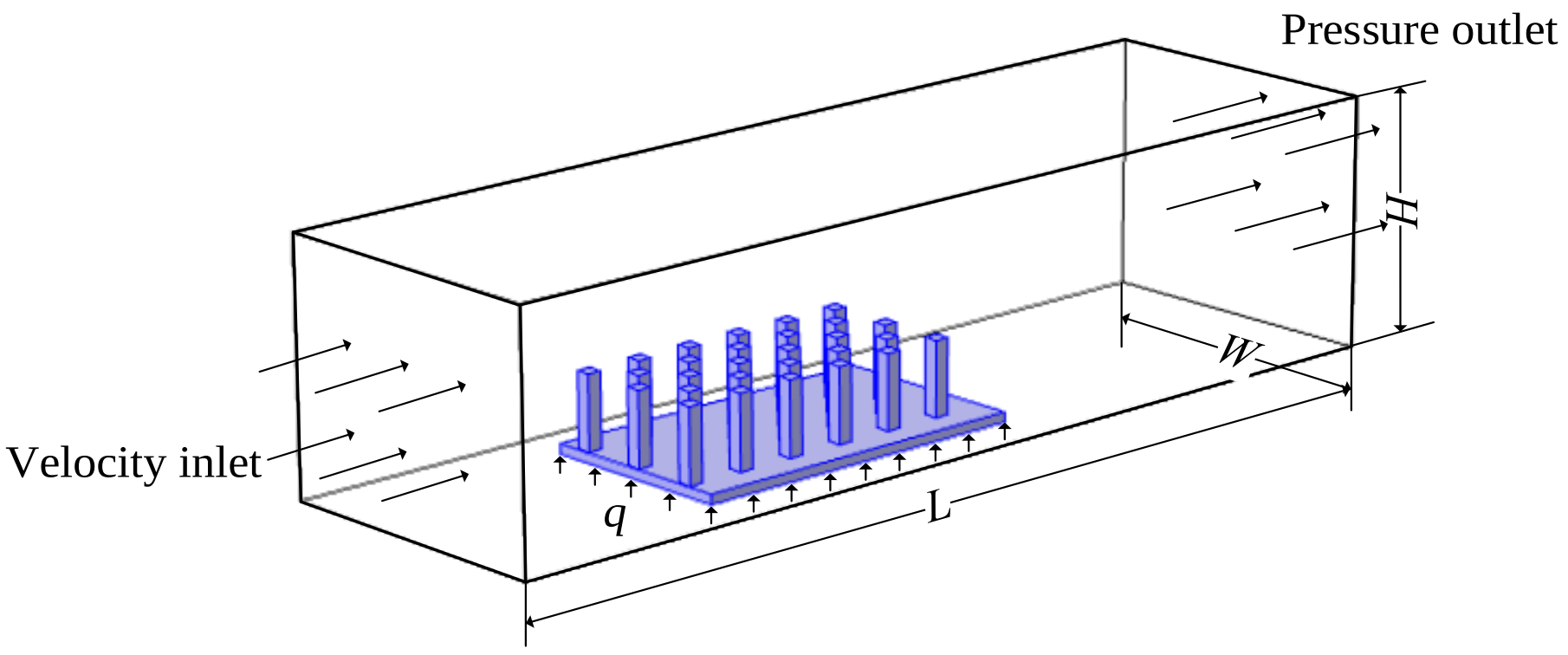

**Fig. 9.** Computational domain.

$$\vec{v} = (u_{in}, 0,0) \qquad T_f = T_{in} \tag{12}$$

where $T_{in} = 25\ ℃$.

The pressure-outlet is applied at the outlet:

$$p = p_{out} \tag{13}$$

where $p_{out} = 0$.

The bottom surface of the heat sink is defined as heat source:

$$-k_s(\partial T_s/\partial n) = q \tag{14}$$

The other sides of computational domain are defined as insulated:

$$\partial T_f/\partial n = 0 \tag{15}$$

Assume good contact between air and aluminum alloy, the interfaces between the two materials should meet the conditions of continuous temperature and heat flux density as follows:

$$T_f = T_s \tag{16}$$

$$k_f(\partial T_f/\partial n) = k_s(\partial T_s/\partial n) \tag{17}$$

According to the analysis result of the PINS as shown in Fig. 10, the average heat flux and average velocity near Face 1 are about 2000 $W/m^2$ and 0.5 *m/s*, respectively. Therefore, the inlet velocity of the FE model is set to 0.5 *m/s*, and a heat flux $q$ = 2000 $W/m^2$ is applied at the bottom of the heat sink.

Moreover, it should be noted that the GSA and the optimization methods used in this study are closed-loop modes. Therefore, the FE model used in this study is parameterized and can be updated adaptive online in each iteration.

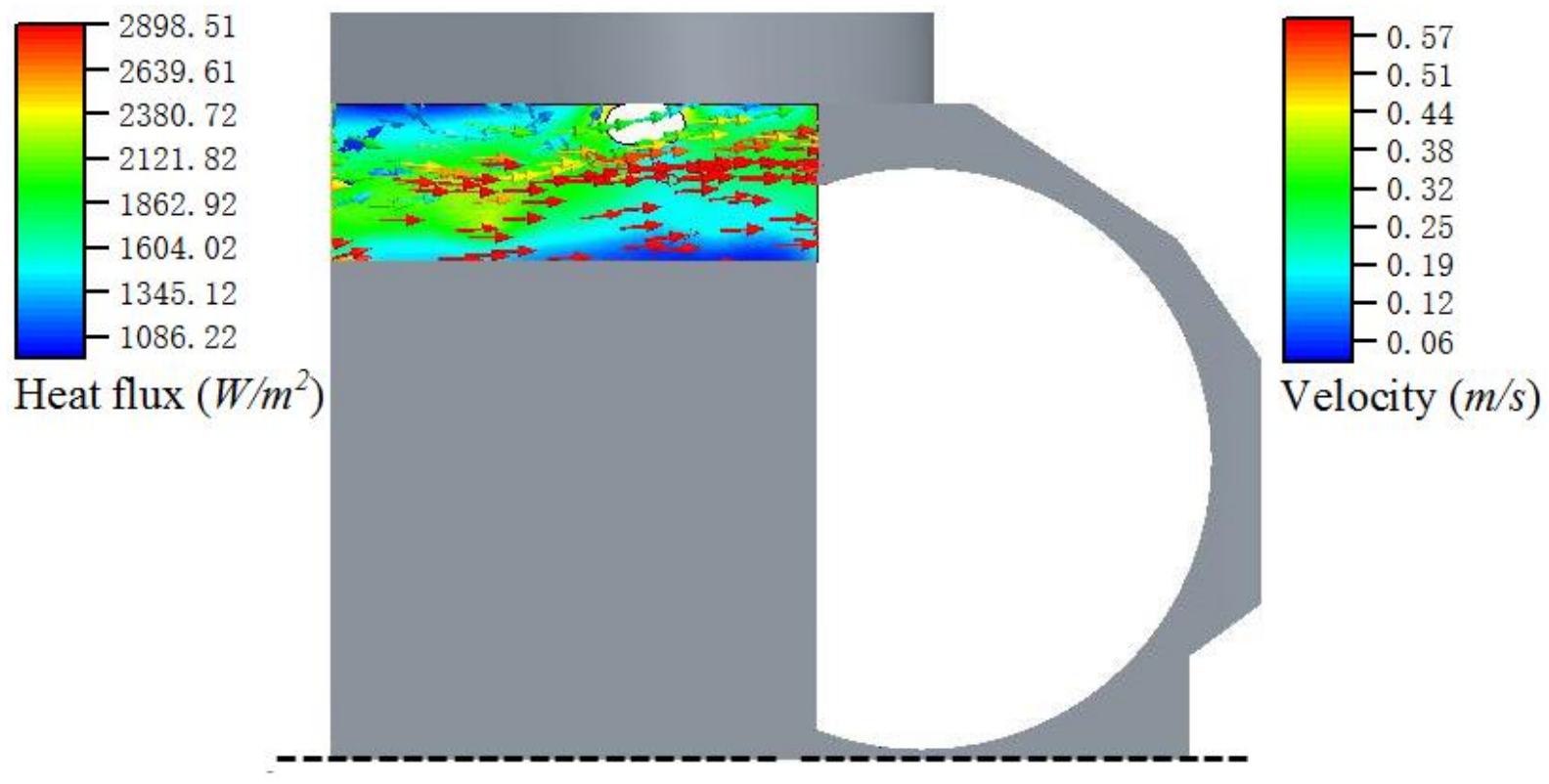

**Fig. 10.** The analysis result of the PINS.

## 4.2. Global sensitivity analysis

The structural parameters $h_f$, $h_d$, $r$, $m_w$, $m_l$, $n$, $\theta$ and the material parameters $\rho$, $C_P$, $k$ are the main factors related to the temperature of the heat sink. These design variables are the inputs in GSA based on DIRECT-KG-HDMR. The initial values of design variables and corresponding bounds are presented in Table 2. Since $m_w$, $m_l$ and $n$ are discrete variables whose design space are shown in Table 2, $m_w$, $m_l$, $n$ sampled by the DIRECT are rounded when performing the sensitivity analysis and optimization design of the heat sink.

**Table 2.** Design variables, initial values and design space.

| Design variables | $h_d$ (*mm*) | $h_f$ (*mm*) | $r$ (*mm*) | $m_l$ - | $m_w$ - | $n$ - | $\theta$ (°) | $\rho$ (*kg/m*$^3$) | $C_P$ (*J*/(*kg K*)) | $k$ (*W*/(*m K*)) |
|---|---|---|---|---|---|---|---|---|---|---|
| Initial value | 2 | 10 | 1.5 | 7 | 3 | 4 | 45 | 2700 | 890 | 155 |
| Lower bound | 1 | 5 | 1 | 2 | 2 | 3 | 0 | 2700 | 233 | 155 |
| Upper bound | 3 | 15 | 2 | 15 | 5 | 10 | 120 | 10530 | 890 | 412 |

According to the assumption that the contact interface temperature is continuous, the maximum temperature in the high temperature region of the platform can be lowered by minimizing the maximum temperature at the bottom of the heat sink. Therefore, the response function was defined as:

$$minimize: \quad f(\mathbf{x}) = T_{\text{max,bottom}} \tag{18}$$

where $T_{\text{max,bottom}}$ is the maximum temperature at the bottom of the heat sink.

Sensitivity indicates the contribution of the design variables to the output response and it is given out based on the Cut-HDMR theory in this study. In order to objectively depict the sensitivities of the design variables, the sensitivity coefficient $s_{x_i}$ is defined as the variation range of the uncoupled term $f_i(x_i)$ calculated by Eq. (7) as:

$$s_{x_i} = f_i(x_i)_{\text{max}} - f_i(x_i)_{\text{min}} \tag{19}$$

Larger value of $s_{x_i}$ indicates the greater contribution of the variable $x_i$ to the response function.

Coupling intensity, which is given out based on the Cut-HDMR theory, indicates the cooperative effects of two variables upon the output response. The coupling coefficient $c_{x_i,x_j}$ is defined as the range of the coupling term $f_{ij}(x_i, x_j)$ calculated by Eq. (7) as:

$$c_{x_i,x_j} = f_{ij}(x_i, x_j)_{\text{max}} - f_{ij}(x_i, x_j)_{\text{min}} \tag{20}$$

Larger value of $c_{x_i,x_j}$ indicates higher coupling intensity between $x_i$ and $x_j$. In this study, $c_{x_i,x_j} < 1.0$ should be screened out and set to be 0 except $c_{k,x_j}$.

The accuracy of the DIRECT-KG-HDMR model is shown in Table 3. The evaluation criteria commonly used in this model are $R^2$, *RAAE*, *RMAE*. $R^2$ is the accuracy criterion to depict the goodness of fit of the model. The maximum value of $R^2$ is 1, bigger $R^2$ is better. *RAAE* is a global criterion to find out the difference between the model forecasts and the actual value produced, smaller *RAAE* is better. *RMAE* is a local criterion that indicates the error of a local region with a relatively poor accuracy of the approximate model, the smaller RMAE indicates better accuracy.

Table 3 shows that the constructed DIRECT-KG-HDMR model has high accuracy and can be used as a surrogate model for the response function in offline optimization.

**Table 3.** The accuracy of the DIRECT-KG-HDMR model.

| Model | Number of samples | $R^2$ | *RAAE* | *RMAE* |
|---|---|---|---|---|
| DIRECT-KG-HDMR | 231 | 0.9834 | 0.1001 | 0.4219 |

The results of the sensitivity analysis are shown in Fig. 11. According to Fig. 11, the diagrams of black dotted frame indicate the result of the sensitivity analysis of the uncoupled terms. The diagrams of red dash-and-dot frame indicate the coupling intensities between structural parameters with large coupling intensity, and the coupling intensities $c_{k,x_j}$ between thermal conductivity $k$ and structural parameters are shown in the diagrams of blue line frame. Because of minimal coupling intensity, the coupling of $\rho$ and $C_P$ with other design variables are neglected.

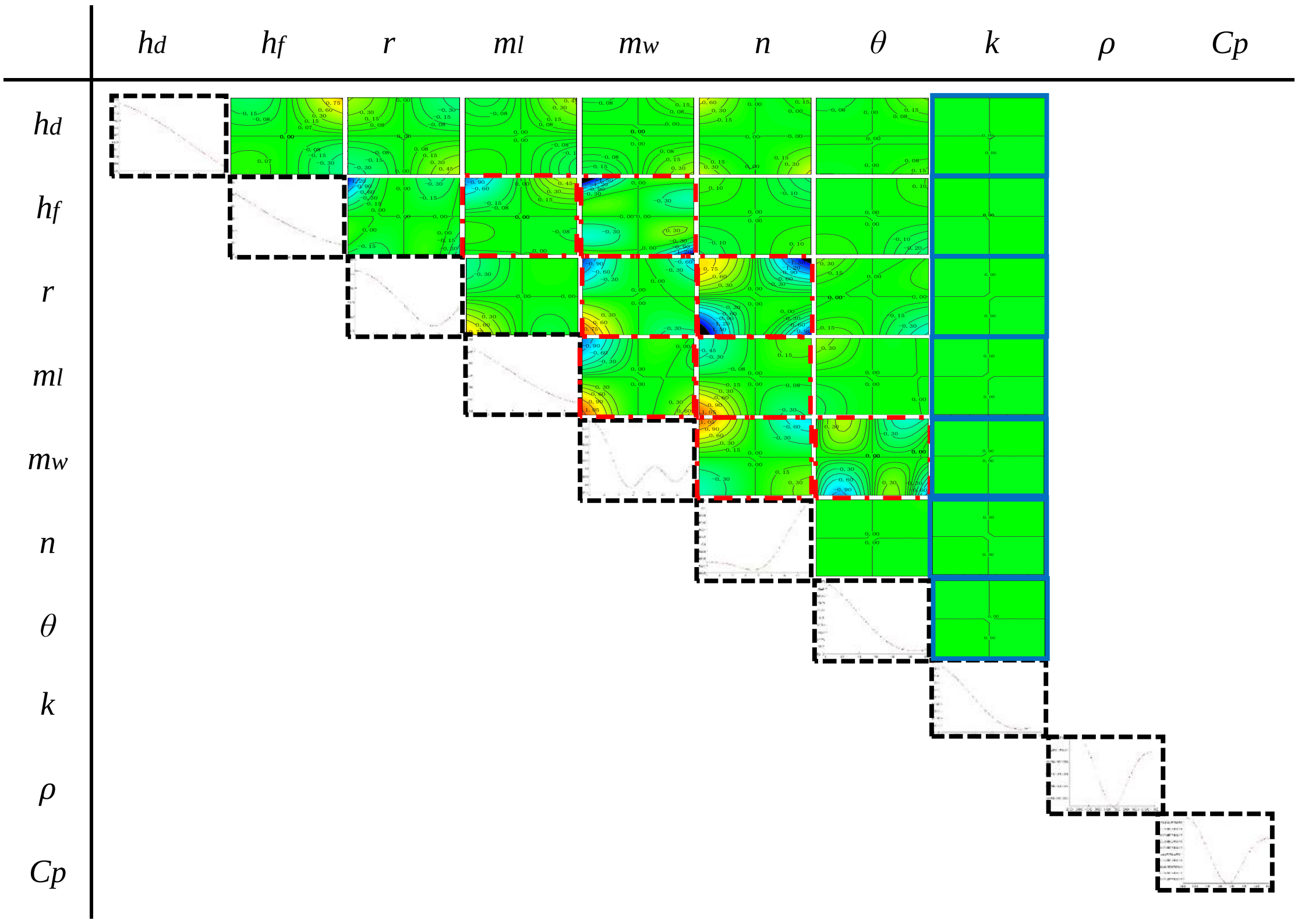

**Fig. 11.** The result of GSA by the suggested method.

The result of GSA of the uncoupled terms after normalization is shown in Fig. 12, where the curves and legend represent the trend and variation of $f_i(x_i)$ in the design interval of $x_i$, respectively. It reveals the effects of the design variables on the maximum temperature at the bottom of the heat sink. As shown in Fig. 12(a), it can be found clearly that the response function is monotonically decreasing in the design interval of $h_d$, $h_f$, $m_w$, respectively. It means that the maximum temperature at the bottom of the heat sink decreases with increasing the design variable, and it is minimized when the design variable $x_i$ reaches the upper bound. It is due to improving heat transfer caused by increasing the heat transfer area. With the increase of $m_l$, the heat transfer area will increase and $d_l$ will decrease. Obviously, heat transfer will be improved with increasing the heat transfer area. On the other hand, reduction of flow velocity caused by decreasing $d_l$ will make heat transfer reduces correspondingly. When $m_l$ is less than 7, the effect of increased heat transfer area on heat transfer is greater than the effect of reducing flow velocity, $f(m_l)$ decreases continuously. Then, $f(m_l)$ fluctuates up and down, it can be inferred that the effect of increased heat transfer area on response function cannot compensate the effect of the reduction of flow velocity. Similar result also can be found from $f(r)$, and $f(r)_{\min}$ is obtained at *r* = 1.75 *mm*. Moreover, the maximum temperature at the bottom of the heat sink reduces firstly and then increased in the design interval of *n*, and $f(n)_{\min}$ is obtained at *r* = 4 or 5.

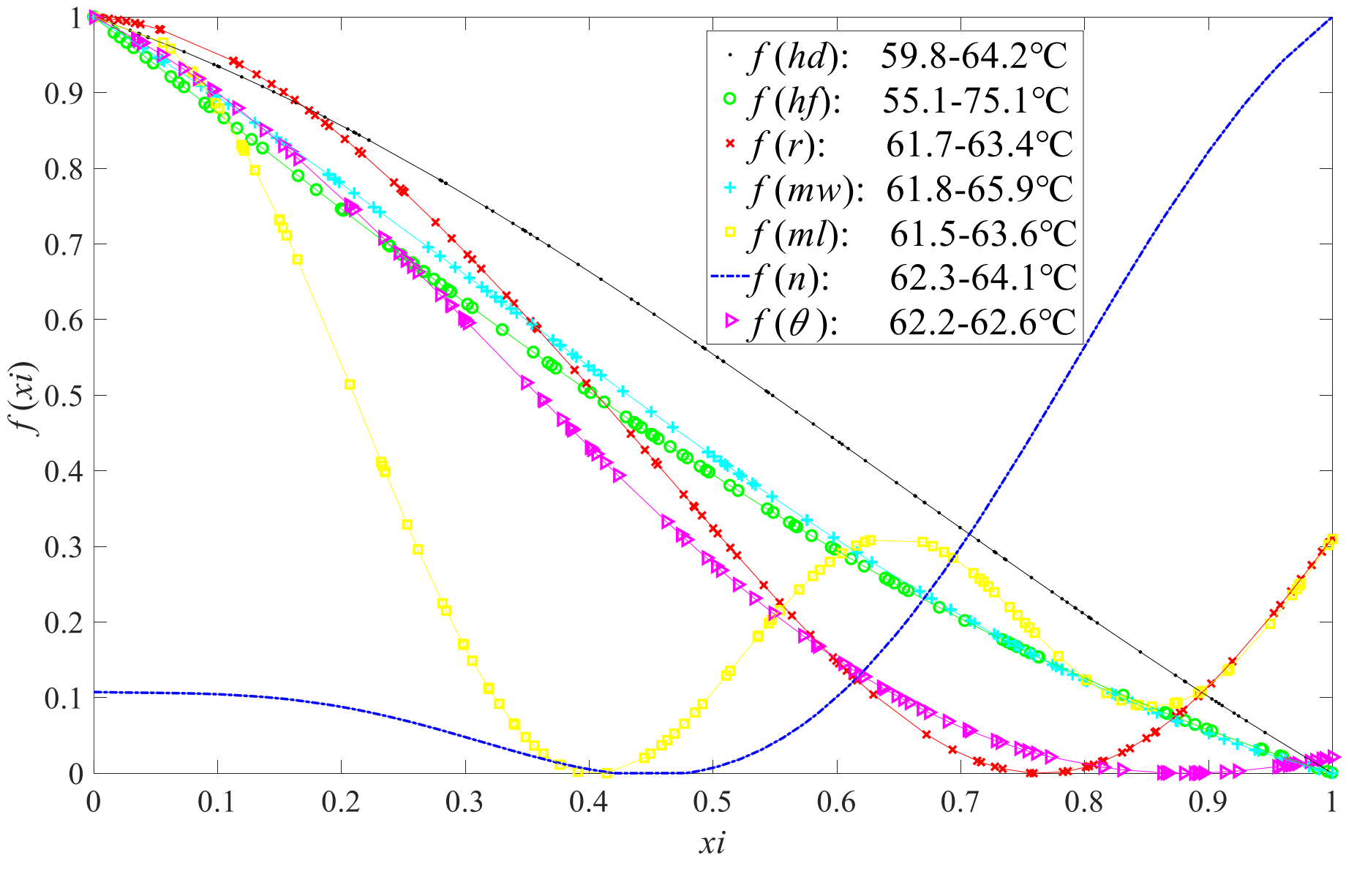

(a) Structural parameters

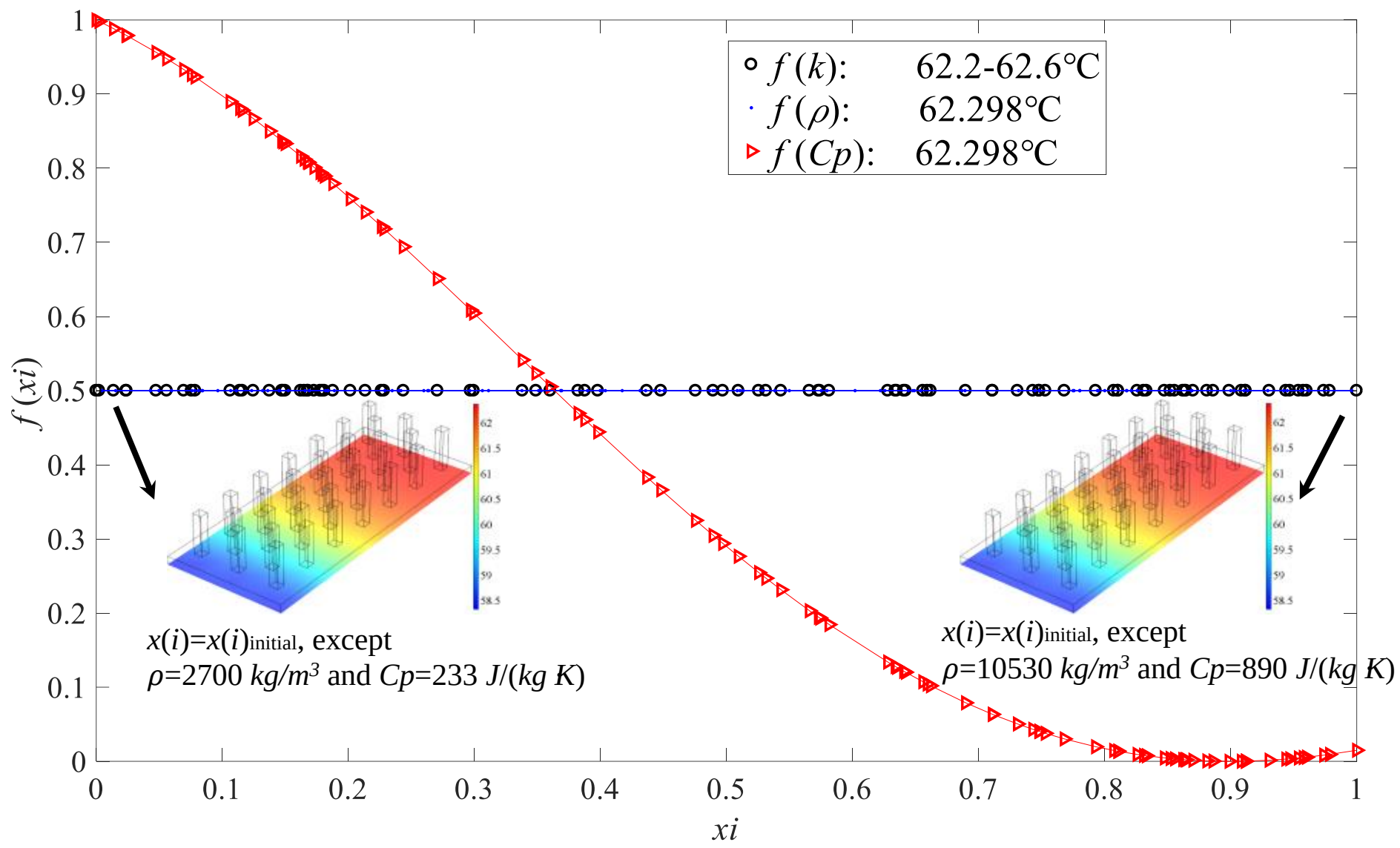

(b) Material parameters

**Fig. 12.** The uncoupled terms of the response function (After normalization).

From Fig. 12, it can be found that $h_f$ is the most important parameter in global variance explanation of the temperature at the bottom of the heat sink, followed by $h_d$, and $m_w$. The remaining input variables such as $n$, $r$ and $m_l$ have much smaller single contributions, and $\theta$ and $k$ are also relatively low. Furthermore, the results of $f(\rho)$ and $f(C_P)$ shown in Fig. 12(b) indicate that the sensitivities of $\rho$ and $C_P$ to the response function are weak, it also can be verified by the temperature contours in Fig. 12(b). Therefore, the influences of $\rho$ and $C_P$ on the response function will not be considered in the following analysis and optimization.

The result of the sensitivity analysis of the coupling terms is shown in Fig. 13 where different colors and contour lines indicate different values of coupling term $f_{ij}(x_i, x_j)$. The coupling intensity between $x_i$ and $x_j$ is represented by the range of colors or the difference between the minimum and maximum values of the contour line. It can be found that $r$ and $m_w$, $r$ and $n$, $m_l$ and $n$, $m_w$ and $m_l$ have the largest coupling intensities among all sets of variances. Moreover, it also can be seen that $h_f$ and $m_l$, $h_f$ and $m_w$, $n$ and $m_w$, $m_w$ and $\theta$ have relatively high coupling intensities. It should be noted that the coupling between $r$ $m_l$ $m_w$ and $n$ is relatively larger, which indicates that the influences of these parameters on the response function are more complicated because they have obvious influences on both flow velocity and heat transfer area. Moreover, the maximum and minimum values of $f_{ij}(x_i, x_j)$ are obtained at the two ends of $x_i$ and $x_j$ for most variable combinations. However, according to the result of $m_w$ and $\theta$, it can be found that the influences of $m_w$ and $\theta$ on $f_{ij}(x_i, x_j)$ are more complicated than the rest. As depicted in Fig. 13(b), $c_{k,x_j} \approx 0$, it means that the coupling intensities between $k$ and other design variables are weak.

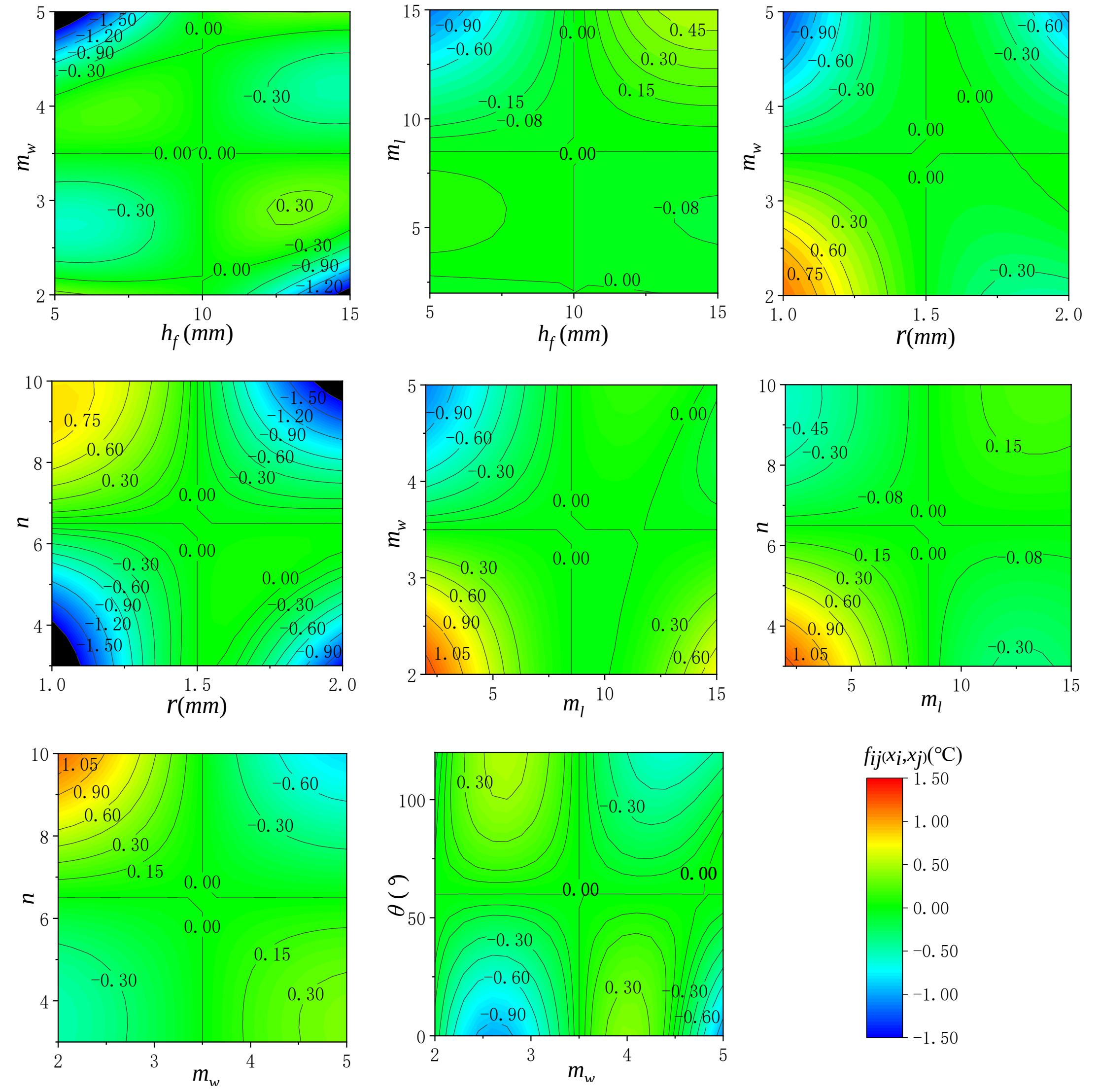

(a)

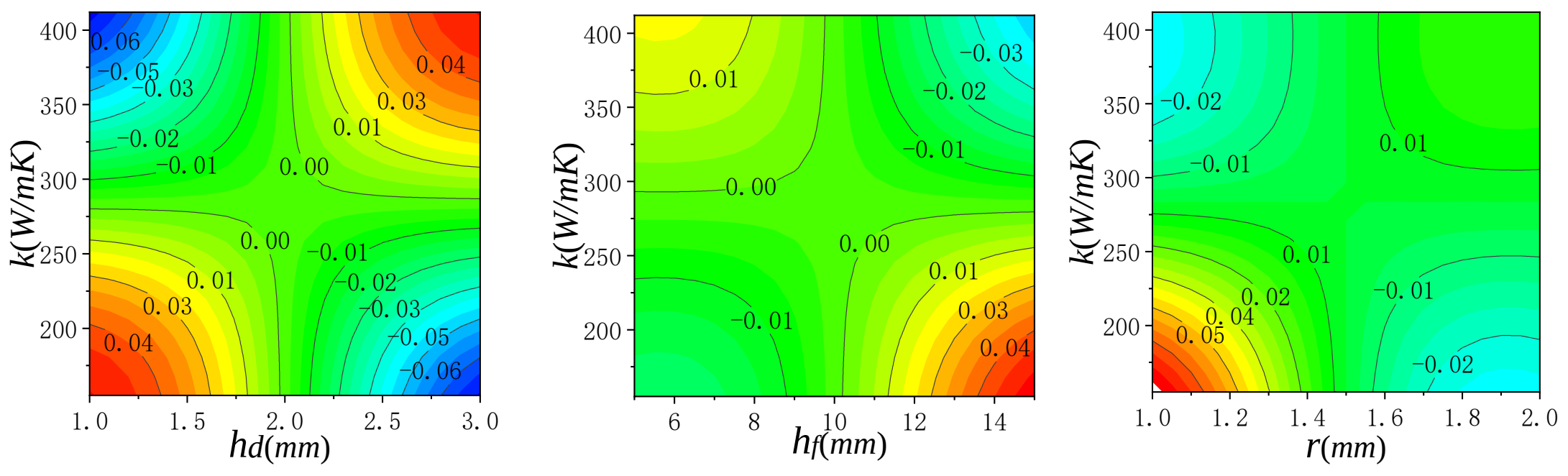

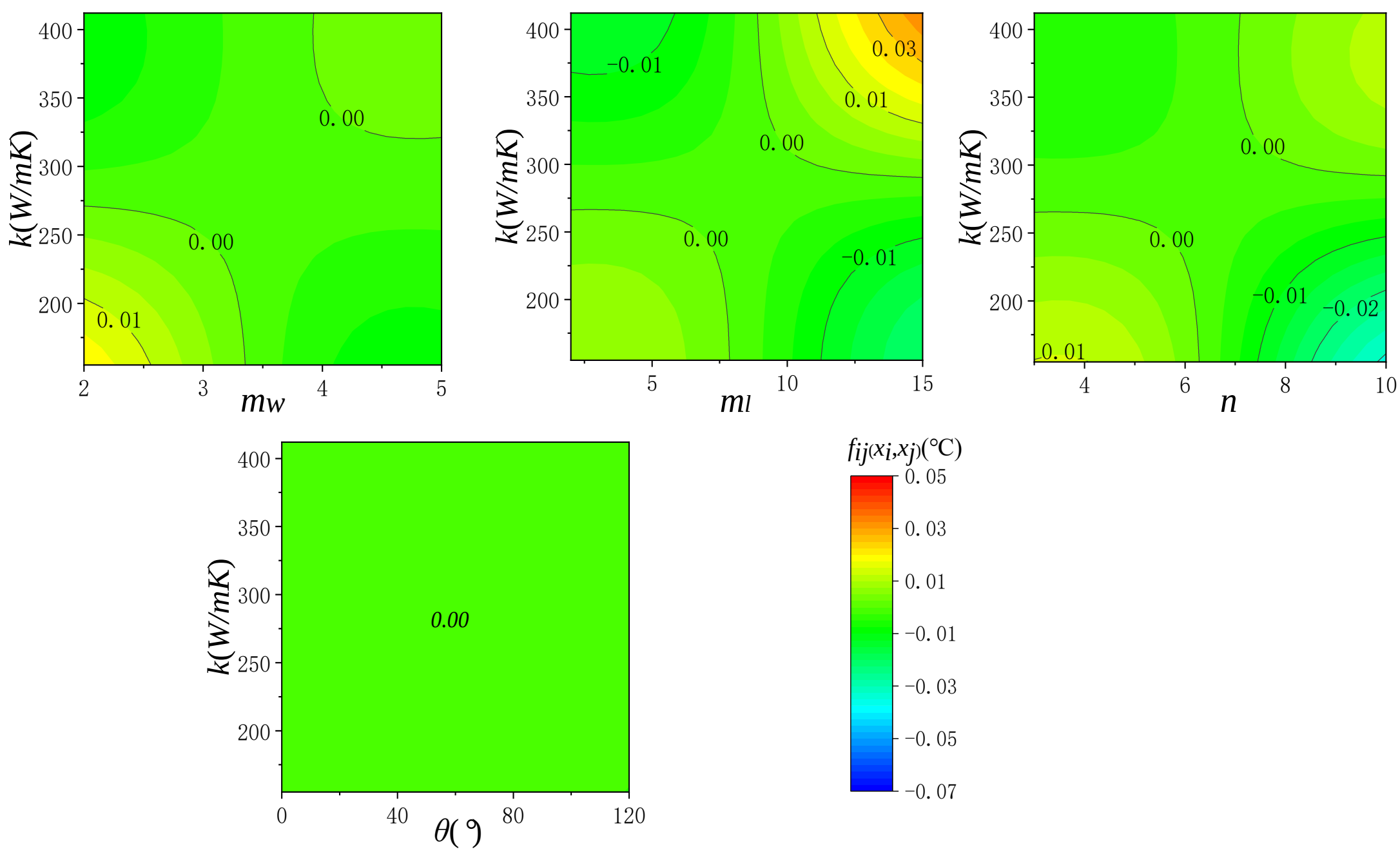

(b)

**Fig. 13.** The coupling terms between: (a) structural parameters; (b) *k* and structural parameters.

According the results mentioned above, it is reasonable to assume that the structural parameters and material parameters can be analyzed independently. It is effective to reduce the dimensions of such problems and improve the computational efficiency in optimization.

## 4.3. Optimization

The structural parameters $h_f$, $h_d$, $r$, $m_w$, $m_l$, $n$, $\theta$ are selected for optimization design according to the result of sensitivity analysis above. The optimization of the heat sink can be mathematically expressed as:

$$\left.\begin{array}{rl} \textit{Minimize:} & f(\mathbf{x}) = T_{\max,\mathrm{bottom}} \\ \textit{Subject to:} & 1\,mm \leq h_d \leq 3\,mm \\ & 5\,mm \leq h_f \leq 15\,mm \\ & 1\,mm \leq r \leq 2\,mm \\ & 0^\circ \leq \theta \leq 120^\circ \\ & m_w \in [2,3,\ldots,5] \\ & m_l \in [2,3,\ldots,15] \\ & n \in [3,4,\ldots,10] \end{array}\right\} \tag{21}$$

Moreover, the structural and material parameters can be analyzed independently based on the result of GSA. Therefore, the heat sink is made of Al6061 because it is the commonly used material for the heat sink.

According to the previous GSA results, it can be found the assigned problem is weak correlated and some of design variables are linear or weak nonlinear with the objective function. Therefore, it can be optimized efficiently by using design variances after GSA. However, to validate our deduction, both Surrogate Assisted Optimization (SAO) and Evolutionary Algorithms (EAs) are carried out in this study. For SAO, Efficient Global Optimization (EGO) algorithm [47] is used; For EAs, Swarm Optimization (PSO), Differential Evolution (DE), Genetic Algorithm (GA) and Teaching Learning Based Optimization (TLBO) are implemented.

**Table 4.** Parameter setting for EAs.

| PSO | | DE | | GA | | TLBO | |
|---|---|---|---|---|---|---|---|
| *Maxit* | 40 | *Maxit* | 40 | *Maxit* | 40 | *Maxit* | 40 |
| *nPop* | 15 | *nPop* | 15 | *nPop* | 15 | *nPop* | 15 |
| $\omega_{min}$ | 0.4 | *F* | 0.8 | *pC* | 0.4 | | |
| $\omega_{max}$ | 1.0 | *pC* | 0.45 | *pM* | 0.8 | | |
| $c_1$ | 1.5 | | | *nC* | 2×round(*pC*×*nPop*/2) | | |
| $c_2$ | 2.0 | | | *nM* | round(*pM*×*nPop*) | | |

The population size (*nPop*) of each algorithm for EAs is set to 15 and the maximum allowed iteration (Maxit) is 40. Other necessary parameters of EAs are listed in Table 4.

**Table 5.** The optimization results by each algorithm.

| Algorithm | Iterative steps | Number of samples | Optimal design variables | Optimal solution |
|---|---|---|---|---|
| GA | 18 | 285 | [2.63,14.29,1.35,5,5,4,30.51] | 52.93℃ |
| DE | 28 | 435 | [3,15,1.52,5,6,3,81.03] | 50.02 ℃ |
| TLBO | 20 | 615 | [2.87,15,1.86,5,5,4,82.51] | 49.83 ℃ |
| PSO | 19 | 300 | [2.80,15,1.70,5,8,4,0] | 49.62 ℃ |
| EGO | - | 433 | [3,15,1.46,5,7,3,77.01] | 49.49 ℃ |

The results of SAO and EAs are shown in Table 5 and the convergence plots are depicted in Fig. 14 and Fig. 15. It can be found that GA converges faster than DE, TLBO, PSO and EGO. However, the optimization results of EGO are better than the other algorithms, followed by PSO and TLBO. It not only verifies the effectiveness of EGO on the weak nonlinear problems, but also reflects the accuracy of KG surrogate model.

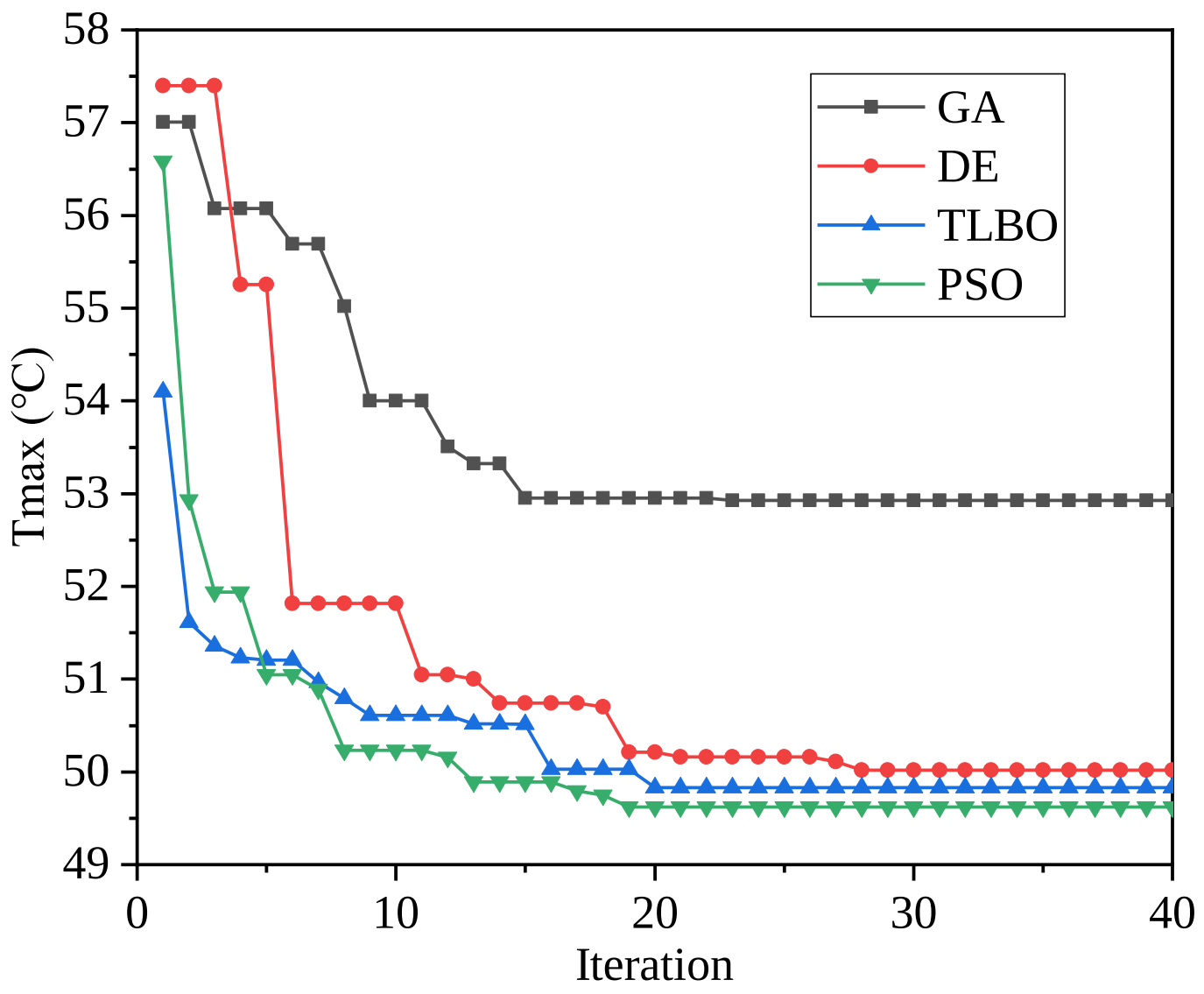

**Fig. 14.** The convergence plots for EAs.

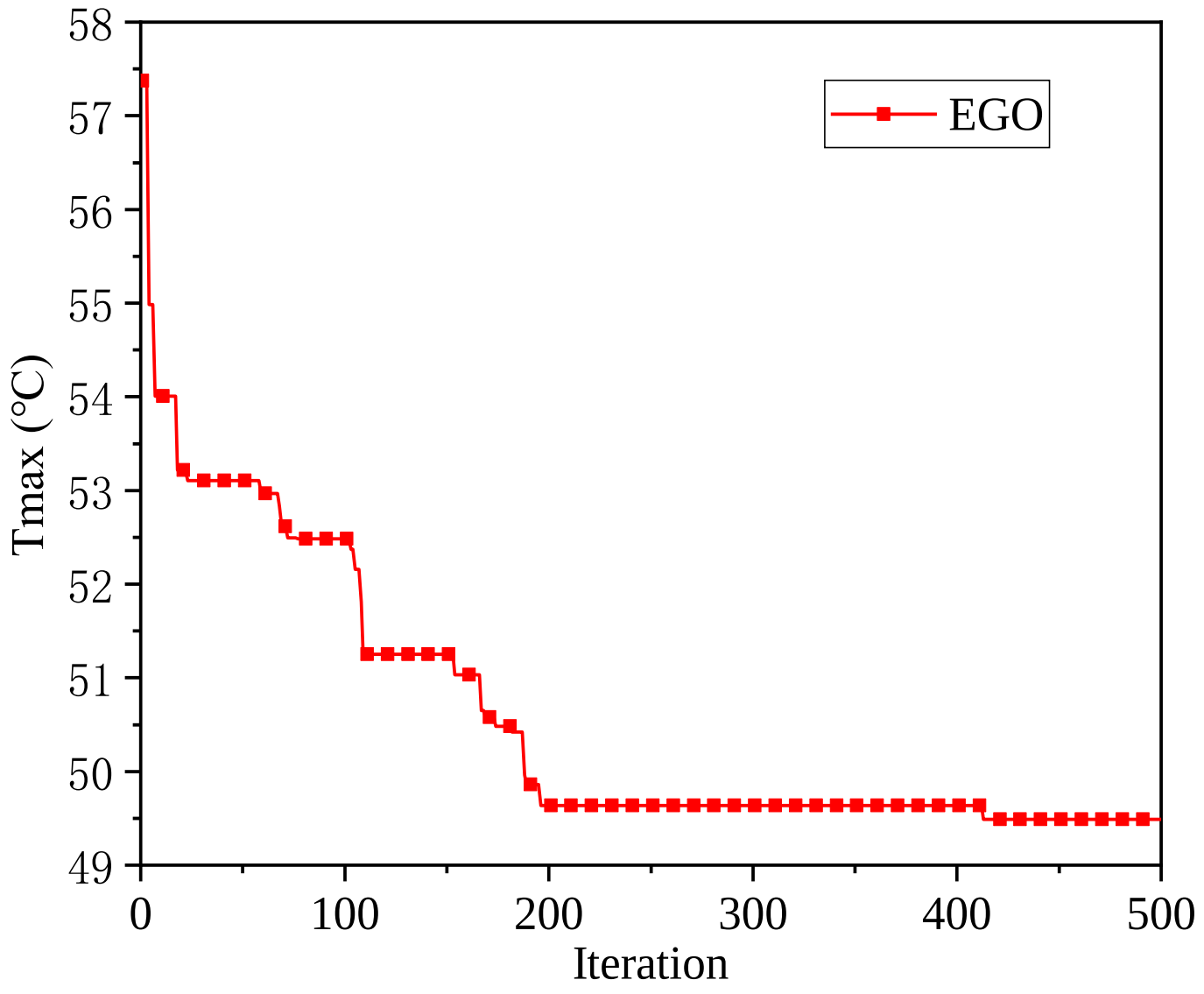

**Fig. 15.** The convergence plot for EGO.

The velocity contours on the planes $x = 15\,mm$ (Mid-Surface) and $z = h_f + h_d$ (Top surface) for the initial and the optimal heat sinks are shown in Fig. 16. The air moves right along the inlet which is due to the inlet velocity of $0.5\,m/s$. When the air starts to contact with the surfaces of the heat sink, the flow velocity increases slightly because of the reduced cross-sectional area of the airflow, the maximum velocity appears in the optimized heat sink by TLBO in Fig. 16(d). It is also noted that the velocity along the left section is higher than the right of heat sink because of decreasing the fin's distance for staggered fins in the flow direction. Staggered fins can disturb the boundary layer and enhance the heat transfer; however, it will make heat transfer reduces because the reduction of flow velocity. Moreover, it can be found in Fig. 16 that different combinations of $n$ and $\theta$ would influence the direction of the air flowing through the heat sink.

The temperature contours at the bottom of the heat sinks are also shown in Fig. 17. It can be observed that the maximum temperature at the bottom of the optimized heat sink is significantly reduced and the temperature of heat sink on the left is less than the right for all cases. The left section is in contact with cold air while the right section is in contact with heated air that moved from the left. It can be found in Fig. 17 that larger $h_f$ can obviously reduce the temperature at the bottom of the heat sink. The value of $n$ is 3 or 4 for four optimized heat sinks, it indicates the agreement of the results of the global sensitivity analysis and optimization. The optimal result shows that best performance of the heat sink appears at $h_d$ = 3 $mm$, $h_f$ = 15 $mm$, $r$ = 1.46 $mm$, $m_w$ = 5, $m_l$ = 7, $n$ = 3, $\theta = 77.01\,°$ and the optimal temperature is 49.49 ℃.

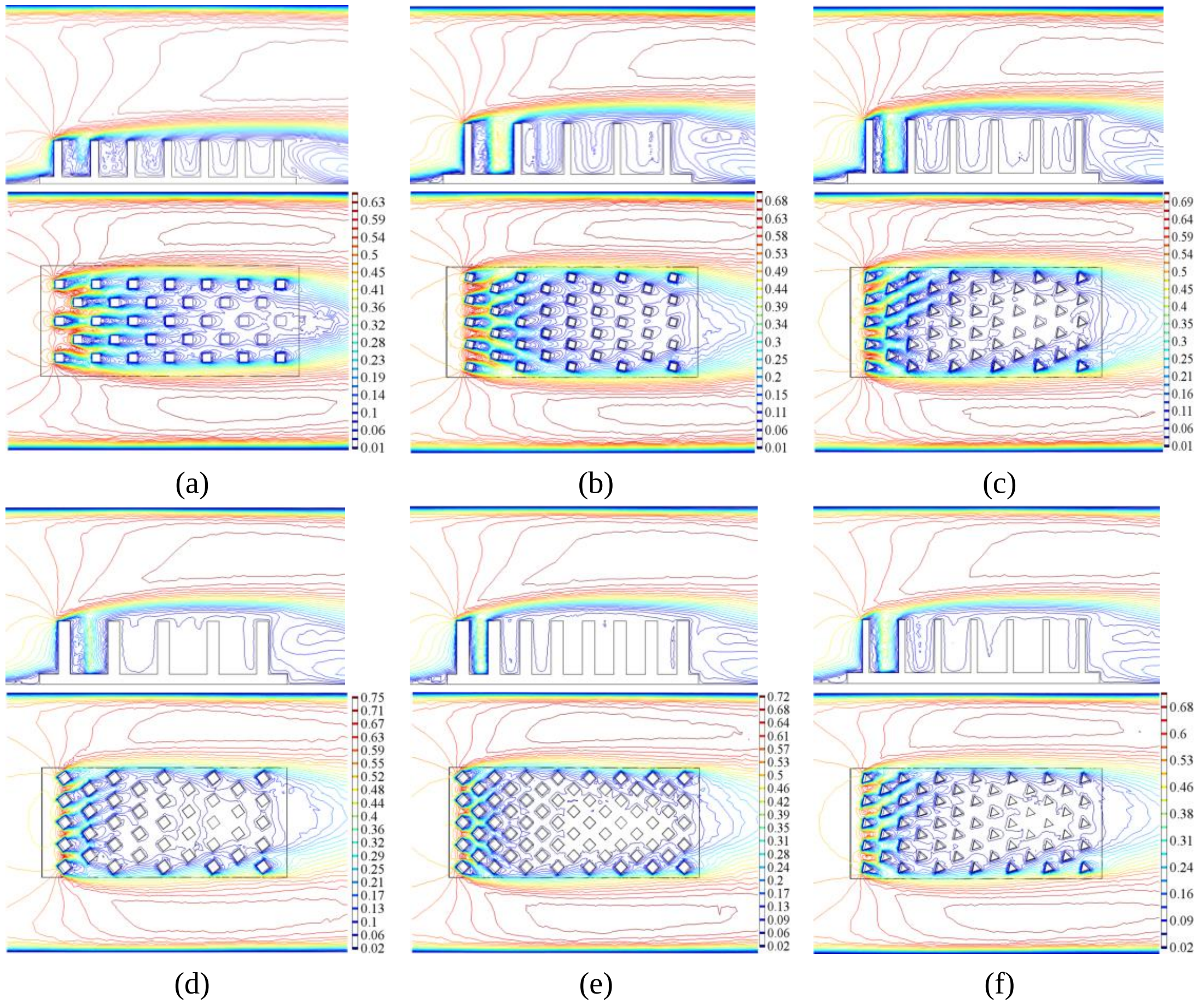

**Fig. 16.** Velocity contours (*m/s*) on the planes $x = 15\,mm$ and $z = h_f + h_d$ for (a) the initial design; (b) the optimized design by GA; (c) the optimized design by DE; (d) the optimized design by TLBO; (e)the optimized design by PSO; (f) the optimized design by EGO.

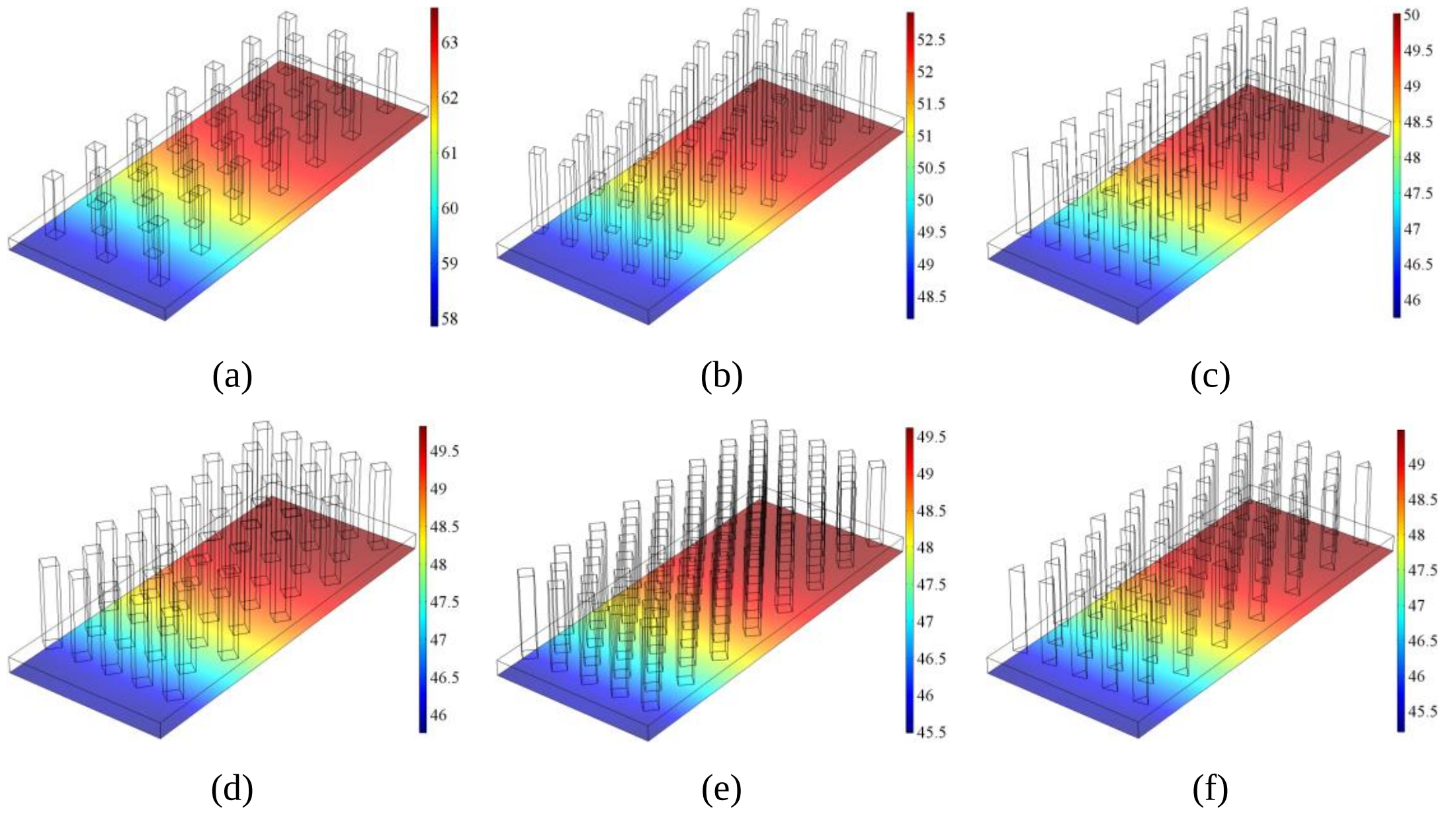

**Fig. 17.** Temperature contours (℃) at the bottom of the heat sink for (a) the initial design; (b) the optimized design by GA; (c) the optimized design by DE; (d) the optimized design by TLBO; (e)the optimized design by PSO; (f) the optimized design by EGO.

## 4.4. Validation of the effectiveness of the heat sink

The temperature contours of Face 1 (as shown in Fig. 2) of three different models are shown in Fig. 18. The maximum temperature of Face 1 is 60.81 ℃ when without heat sink, it drops by 1.09 ℃ with the optimized heat sink. The drop of the maximum temperature of the heat sink seems not obvious in magnitude. However, the accuracy of PINS is extremely sensitive to temperature, imprecise measurements caused by tiny temperature rise even 0.1℃ would lead to disastrous accident, even tiny temperature reduction would improve the precision of PINS. Therefore, the performance of the platform has been improved well in this work.

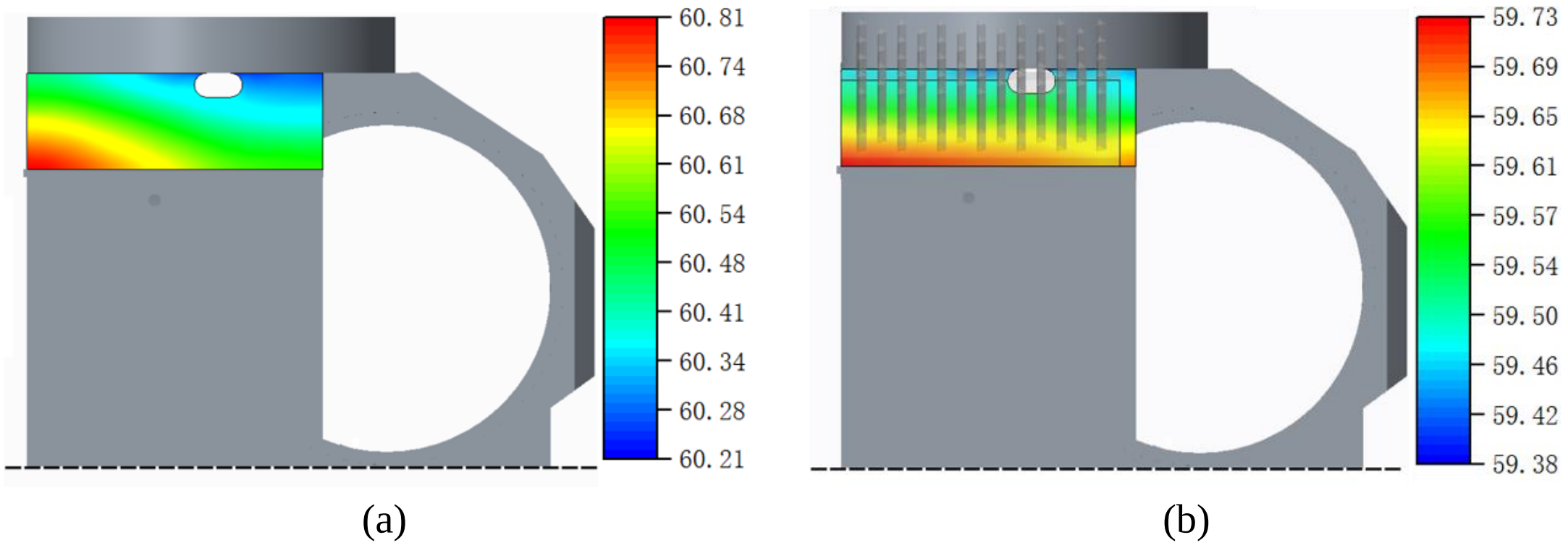

**Fig. 18.** The temperature contours (℃) for (a) Without heat sink; (b) With the optimized heat sink by EGO.

# 5. Conclusion

In this study, a heat sink using pin fins with staggered arrangement is designed to reduce the local high temperature of the platform. A DIRECT-KG-HDMR based sensitivity analysis of the structural parameters $h_f$, $h_d$, $r$, $m_w$, $m_l$, $n$, $\theta$ and material parameters $\rho$, $C_P$, $k$ on the maximum temperature at the bottom of the heat sink is presented. Then, sensitivity coefficient $s_{x_i}$ and coupling coefficient $c_{x_i,x_j}$ are defined to indicate the effects of design variables on the response function. The result shows that $h_f$ is the most sensitive input on the response function, followed by $h_d$, $m_w$, $m_l$, and it is also validated that $\rho$ and $C_P$ have almost no influence on $f(x)$. Furthermore, it is shown that $r$ and $m_w$, $r$ and $n$, $m_l$ and $n$, $m_l$ and $m_w$ have the largest coupling intensities among all pairs of variances. However, the coupling intensities between $k$ and other design variables are weak, it is reasonable to infer that the structural parameters and material parameters can be analyzed independently for optimization design. According to the result of sensitivity analysis, the structural parameters $h_f$, $h_d$, $r$, $m_w$, $m_l$, $n$, $\theta$ are selected for optimization, and the optimization results and efficiency of GA, DE, TLBO, PSO and EGO were compared. The result shows that GA converged faster than others. However, the best optimal solution is obtained by EGO. Moreover, it is found that the response function is minimal when the design variable $h_d$, $h_f$, $m_w$ reach the upper bound at the same time, and it is consistent with the results of GSA. Finally, PINS with the optimized heat sink by EGO is simulated and the result shows that the maximum temperature of high temperature region of the platform reduces when the optimized heat sink is mounted on the platform.

# Acknowledgments

This work has been supported by Project of the Key Program of National Natural Science Foundation of China under the Grant Numbers 11572120 and 51621004.

# Appendix

## *A.1. Particle Swarm Optimization (PSO)*

The PSO algorithm, which was proposed by Kennedy and Eberhart [42], is a global random search algorithm based on swarm intelligence, and it is inspired by social interaction of animals living in groups. An initial particle swarm is generated randomly firstly, and it is also the candidate solution of the problem. Then, the fitness value $F_{ij}(i)$ of each particle is calculated and compared with the best point $p_{best}(i)$ of the particle, if $F_{ij}(i) > p_{best}(i)$, replacing $p_{best}(i)$ with $F_{ij}(i)$. Then, $F_{ij}(i)$ is compared with the best point $g_{best}$ among the swarm, if $F_{ij}(i) > g_{best}(i)$, replacing $g_{best}(i)$ with $F_{ij}(i)$. The process terminates when the optimal solution is obtained by continuous iteration or reaches the maximum iteration.

It should be noted that each particle is characterized by two main properties: position and velocity. The position of particle is the evaluation of the optimization's fitness function, which represents the particles distance from the ideal optimum solution. Similarly, the velocity of particle dictates the motion of the particle in the solution space. During the iteration $t + 1$, the position of each particle is updated by:

$$\mathbf{x}_i(t+1) = \mathbf{x}_i(t) + \mathbf{v}_i(t+1) \tag{22}$$

and its velocity is updated by :

$$\begin{aligned}\mathbf{v}_i(t+1) = \omega\mathbf{x}_i(t) + c_1 \cdot rand(0,1) \cdot \left(\mathbf{x}_{li}(t) - \mathbf{x}_i(t)\right) \\ + c_2 \cdot rand(0,1) \cdot \left(\mathbf{x}_g(t) - \mathbf{x}_i(t)\right)\end{aligned} \tag{23}$$

where $c_1$ and $c_2$ are the personal learning coefficient and global learning coefficient, respectively. $\omega$ is the inertia factor. The larger $\omega$ is beneficial to jump out of the local minimum point, it is convenient for global search. The smaller $\omega$ is beneficial to the accurate local search, which is conducive to algorithm convergence. Therefore, in order to balance the efficiency and accuracy, a linear function of $\omega$ is applied:

$$\omega = \omega_{min} + \frac{(Maxit - t) \cdot (\omega_{max} - \omega_{min})}{Maxit} \tag{24}$$

where $\omega_{max}$ and $\omega_{min}$ are the maximum and minimum of $\omega$, respectively. $t$ and $Maxit$

represent the current iteration and the maximum iteration number, respectively.

## *A.2. Differential Evolution (DE)*

The DE algorithm is a heuristic random search algorithm based on group difference, which was proposed by Storn and Price [43]. Compared with the evolutionary algorithm, DE retains the population-based global search strategy, and it reduces the complexity of genetic operations by adapting one-to-one competitive survival strategy and simple mutation operation. Because of unique memory ability, it is possible for DE to track the current search situation dynamically and adjust its search strategy when needed. Moreover, it is suitable for solving the optimization problems that cannot be solved using conventional mathematical programming methods in complex engineering. The basic idea of DE is: (1) Mutation. The *m*-dimensional difference vector of two individuals randomly selected from the population is used as the random mutation source of the third individual, then the variant individual is generated by summing the weighted difference vector and the third individual. The mutation individual can be expressed as:

$$\mathbf{v}_i(j+1) = \mathbf{x}_{r1}(j) + F \cdot \left(\mathbf{x}_{r2}(j) - \mathbf{x}_{r3}(j)\right) \qquad i \neq r_1 \neq r_2 \neq r_3 \tag{25}$$

where *F* is the scaling factor, and $\mathbf{x}_i(j)$ is the *no. i* individual of *no. j* population.

(2) Crossover. The crossover between $\mathbf{x}_i(j)$ and $\mathbf{v}_i(j+1)$ can be expressed as:

$$\begin{aligned} \mathbf{u}_{ki}(j+1) &= \mathbf{v}_{ki}(j+1), \quad if\ rand(0,1) \leq\ pC\ or\ k = k_{rand} \\ &= \mathbf{x}_{ki}(j). \qquad otherwise \end{aligned} \tag{26}$$

Where *pC* is the crossover rate, and $k_{rand}$ is a random integer from 1 to *m*.

(3) Selection. Greedy algorithm is applied to select the individuals of the next population:

$$\begin{aligned} \mathbf{x}_i(j+1) &= \mathbf{u}_i(j+1), \quad if\ rand(0,1) \leq\ pCR\ or\ k = k_{rand} \\ &= \mathbf{x}_i(j). \qquad otherwise \end{aligned} \tag{27}$$

The choice of DE parameters *F* and *pC* can have a large impact on optimization performance. According to the rules for parameter selection [44], *F* and *pC* are set to 0.8 and 0.45, respectively.

## *A.3. Genetic Algorithm (GA)*

GA is an optimal algorithm to search for optimal solutions by simulating the process of biological evolution [45]. It begins with a population that represents a potential set of solutions to a problem, and it consists of a certain number of individuals encoded by genes. Each individual is actually an entity with a characteristic chromosome. Chromosome is the main carrier of genetic material and its internal representation is a combination of genes that determines the external representation of individual shape. Therefore, what should be done is coding work and mapping from phenotype to genotype at the beginning. After the initial generation of the population, evolution of generations produces better approximate solutions based on the principle of survival of the fittest. At each iteration, each individual is selected according to the fitness of the individual in the design space, and the crossover and mutation are altered by adaptive probabilities to generate a new population. At last, the best individual in the last generation is decoded, which can be used as the approximate optimal solution of the engineering problem.

## *A.4.Teaching-Learning-Based Optimization (TLBO)*

The TLBO algorithm is a method inspired by teaching-learning process [46]. A series of solutions in the design space are randomly generated, they are regarded as the students in different subjects. There are two basic modes of learning in the class: (i) the process of learning from teacher (known as teacher phase), and (ii) the process of learning with the other students (known as student phase). In this algorithm, a group of students is considered as population and the different subjects offered to the students are considered as different design variables of the optimization problem and a student's score is analogous to the fitness value of the optimization problem. The best solution in the entire population is considered as the teacher. After repeated teacher phase and student phase, the level of knowledge of the class is getting higher and higher, which means that the search space of the feasible solution approaches the optimal solution in the whole restricted space.

Firstly, the students' scores $\mathbf{x}^j = \left(x_1^j, x_2^j, \cdots x_d^j\right)(j = 1,2,\cdots,nPop)$ are initiated randomly in the restricted space.

$$\mathbf{x}_i^j = \mathbf{x}_i^L + \text{rand }(0,1) \cdot (\mathbf{x}_i^U - \mathbf{x}_i^L) \tag{28}$$

Then, in the teacher phase of TLBO, every student of the class will learn based on the *difference* between the teacher's scores $\mathbf{x}_{teacher}$ and the mean of the students' scores $\mathbf{x}_{mean}$. After repeated teacher phase, the mean of the class gradually increased from the lower $\mathbf{x}_{mean,A}$ to the higher $\mathbf{x}_{mean,B}$, and the distribution of the scores became relatively concentrated. The process can be expressed as

$$\mathbf{x}_{new}^{i} = \mathbf{x}_{old}^{i} + difference \tag{29}$$

$$difference = r_i + (\mathbf{x}_{teacher} - TF_i \cdot \mathbf{x}_{mean}) \tag{30}$$

$$\mathbf{x}_{mean} = \frac{1}{nPop}\sum_{i=1}^{nPop} x^{i} \tag{31}$$

where $\mathbf{x}_{old}^{i}$ and $\mathbf{x}_{new}^{i}$ are the scores of *no.i* student before and after teacher phase, respectively. $TF_i = \mathrm{round}[1 + \mathrm{rand}(0,1)]$ is the teaching factor and $r_i = \mathrm{rand}(0,1)$ is the learning factor.

The scores will update after every teacher phase, if $f(\mathbf{x}_{new}^{i}) > f(\mathbf{x}_{old}^{i})$, $\mathbf{x}_{old}^{i} = \mathbf{x}_{new}^{i}$. The process terminates when the optimal solution is obtained by continuous iteration or reaches the maximum iteration.

## *A.5. Efficient Global Optimization (EGO)*

The EGO algorithm, which is proposed by Jones et al. [47], is effective for both local search and global search. It is especially suitable for the weak nonlinear problems with fewer parameters. Firstly, the initial sample points are sampled by Latin Hypercube Sampling (LHS) and the number of initial sample points is $k = 3 \cdot n$, where *n* denotes the number of input variables. A KG surrogate model is constructed based on the initial samples. Then, a new sample point is added to the sample set based on add-point function. The function of improving the expectation is most commonly used as add-point function because it is a method to weight the predicted value and the predicted variance, and it would calculate the probability of getting a better response at a given point. For example, the response $\hat{y}(\mathbf{x})$ and the standard deviation $\sigma(\mathbf{x})$ are predicted by KG model, the improvement for a sample $\mathbf{x}$ can be expressed as:

$$I(\mathbf{x}) = \max(y_{\min} - \hat{y}(\mathbf{x}), 0) \tag{32}$$

where $y_{\min}$ is the present optimum per cycle. The expectation of $I(\mathbf{x})$ can be expressed as:

$$EI(\mathbf{x}) = \left(y_{\min} - \hat{y}(\mathbf{x})\right) \times \Phi\left(\frac{y_{\min} - \hat{y}(\mathbf{x})}{\sigma(\mathbf{x})}\right) + \sigma(\mathbf{x}) \times \varphi(\frac{y_{\min} - \hat{y}(\mathbf{x})}{\sigma(\mathbf{x})}) \tag{33}$$

where $\Phi(\cdot)$ and $\varphi(\cdot)$ are the cumulative density function (CDF) and the probability density function (PDF) of a normal distribution, respectively. The process terminates when the optimal solution is obtained by continuous iteration or reaches the maximum iteration.